\documentclass[preprint,12pt,authornum]{elsarticle}

\usepackage{multirow}
\usepackage{amsmath}
\usepackage{amssymb}
\usepackage{graphicx}
\usepackage{subcaption}
\usepackage{caption}
\usepackage[utf8]{inputenc}
\usepackage{geometry}
\usepackage{xcolor}
\usepackage{enumerate}
\usepackage{enumitem}
\usepackage[ruled, vlined]{algorithm2e}
\usepackage{algorithmic}
\usepackage{setspace}
 \usepackage{url}
\usepackage[colorlinks = true,
            linkcolor = blue,
            urlcolor  = blue,
            citecolor = blue,
            anchorcolor = blue]{hyperref}
\usepackage{natbib}
\usepackage{booktabs} 

\journal{arXiv}

\begin{document}

\begin{frontmatter}




\title{Surrogate modeling for probability distribution estimation: uniform or adaptive design?}


\author[inst1]{Maijia Su\corref{cor2}}
\author[inst2]{Ziqi Wang}
\author[inst1]{Oreste Salvatore Bursi}
\author[inst1]{Marco Broccardo\corref{cor1}}

\affiliation[inst1]{organization={Department of Civil, environmental and mechanical engineering, University of Trento, Trento},
            country={Italy}}
\affiliation[inst2]{organization={Department of Civil and Environmental Engineering, University of California, Berkeley},
            country={United States}}

\cortext[cor1]{marco.broccardo@unitn.it}
\cortext[cor2]{maijia.su@unitn.it}

\begin{abstract}
The active learning (AL) technique, one of the state-of-the-art methods for constructing surrogate models, has shown high accuracy and efficiency in forward uncertainty quantification (UQ)  analysis. This paper provides a comprehensive study on AL-based global surrogates for computing the full distribution function, i.e.,  the cumulative distribution function (CDF) and the complementary CDF (CCDF). To this end, we investigate the three essential components for building surrogates, i.e.,  types of surrogate models, enrichment methods for experimental designs, and stopping criteria. For each component, we choose several representative methods and study their desirable configurations. In addition, we devise a uniform design (i.e, space-filling design) as a baseline for measuring the improvement of using AL. Combining all the representative methods, a total of 1,920 UQ analyses are carried out to solve 16 benchmark examples. The performance of the selected strategies is evaluated based on accuracy and efficiency. In the context of full distribution estimation, this study concludes that (\textit{i}) AL techniques cannot provide a systematic improvement compared with uniform designs, (\textit{ii}) the recommended surrogate modeling methods depend on the features of the problems (especially the local nonlinearity), target accuracy, and computational budget.
\end{abstract}

\begin{keyword}
Uncertainty quantification \sep Full distribution function \sep Surrogate models \sep Active learning \sep Space-filling design
\end{keyword}



\end{frontmatter}


\section{Introduction}
\label{sec1}
\noindent
Real-world engineering systems inevitably involve uncertainties, which may arise from physical properties (e.g., material strength and geometric dimensions) and ambient conditions (e.g., external load and environmental noises). These uncertainties are often represented by probability distributions that describe the likelihood of certain variables occurring within a given range. Uncertainty Quantification (UQ) is a field of study that aims to understand how these uncertainties affect the overall performance of a system. For this purpose, UQ techniques are used to propagate the randomness of the basic variables and determine the statistical properties of the system responses. The results of UQ can be used to inform safety assessments, optimize designs, and make risk-informed decisions. This study focuses on a key aspect of the UQ analysis: estimating the full probability distribution function, including the tails (up to prescribed fractiles of interest) of both the cumulative distribution function (CDF) and the complementary CDF (CCDF).

In this context, the source of uncertainty is modeled by a random vector denoted by $\boldsymbol{X} = [X_1,X_2,...,X_N]$ with a prescribed joint probability density function (PDF)  $f_{\boldsymbol{X}}(\boldsymbol{x})$. The system output $Y$, propagated from $\boldsymbol{X}$, represents the Quantity of Interest (QoI) selected to describe the system performance. In general, the QoI can be a multivariate output, but this work only focuses on the unidimensional case. The connection between $\boldsymbol{X}$ and $Y$ is built by a deterministic simulator $\mathcal{M}:\boldsymbol{x}\in\mathbb{R}^N\mapsto y\in\mathbb{R}$. In practice, evaluating $\mathcal{M}(\cdot)$ is computationally expensive since it may involve complex numerical models, e.g., finite element models (FEMs) and computational fluid dynamics models. Given these definitions, the CDF $F_Y(y)$ and CCDF $\bar{F}_Y(y)$ can be expressed as \cite{Wang2020}:
\begin{equation}
 \begin{aligned}
\label{sec1_eq1}
  F_Y(y) &= \mathbb{P}(Y\leq{y})=\mathbb{P}(\mathcal{M}(\boldsymbol{X})\leq{y}) \\
                                    & = \int_{\mathcal{M}(\boldsymbol{x})\leq{y}}f_{\boldsymbol{X}}(\boldsymbol{x})d\boldsymbol{x}
\end{aligned},
\end{equation}
and 
\begin{equation} 
\begin{aligned}
\label{sec1_eq2}
  \bar{F}_Y(y) &= \mathbb{P}(Y>{y})=\mathbb{P}(\mathcal{M(\boldsymbol{X})}>{y}) \\
                 & = \int_{\mathcal{M}(\boldsymbol{x})>{y}}f_{\boldsymbol{X}}(\boldsymbol{x} )d\boldsymbol{x} 
                   = 1-F_Y(y)               
\end{aligned}.
\end{equation}
Despite the trivial relationship between Eqs.\eqref{sec1_eq1} and \eqref{sec1_eq2}, these equations need to be evaluated separately. This is because the low probability regions of CDF and CCDF span opposite tails of the PDF $f_Y(y)$. The separate computations ensure the accurate estimation of the full distribution, including the low-probability regions.

To solve Eqs.\eqref{sec1_eq1} and \eqref{sec1_eq2}, the distribution function can be discretized into a set of thresholds $[{{y}_{\min }}<...<{{y}_{m}}<...<{{y}_{\max }}]$.  Then, the full distribution can be approximated at a series of positions $y_m$, i.e., to compute a sequence of multiple integrals of $f_{\boldsymbol{X}}(\boldsymbol{x})$ on the region ${\mathcal{D}}_{m}= \{ \boldsymbol{x}\in {\mathbb{R}}^{N} | \mathcal{M}(\boldsymbol{x}) \le {y}_{m} \}$ for $F_Y(y_m)$ or on the region  $\bar{{\mathcal{D}}}_{m}= \{ \boldsymbol{x}\in {\mathbb{R}}^{N} | \mathcal{M}(\boldsymbol{x}) > {y}_{m} \}$ for $\bar{F}_Y(y_m)$. Therefore, the UQ problems are converted into common integration problems, which can be solved with a wide variety of methods. Recently, non-intrusive methods receive wide attention since $\mathcal{M}(\cdot)$ can be taken as a black-box simulator, allowing the UQ analysis to be separated from solving the deterministic equations governing the problem at hand. According to \cite{Lee2008}, these approaches can be categorized as:  (\textit{i}) deterministic numerical integration methods (e.g., classical numerical integration \cite{Evans1972}), (\textit{ii}) stochastic simulation methods (e.g., Monte Carlo simulation \cite{Robert2004}, importance sampling \cite{Melchers1989}, subset simulation \cite{Au2001}), (\textit{iii}) local approximation methods (e.g.,  FORM and SORM \cite{ditlevsen1996structural}, Taylor series method \cite{madsen2006methods}) and (\textit{iv}) surrogate-model aided methods (see an overview in \cite{sudret2017surrogate}). In practice, the best approach strongly relies on the target accuracy, computational budget, dimensionality, and the specific features of the problem. 

In recent decades, surrogate modeling methods have gained popularity since they can save substantial computational costs while providing relatively high-precision estimations.
Surrogate models, a.k.a. metamodels, seek to find an inexpensive approximation model $\hat{\mathcal{M}}(\cdot)$ to substitute the original simulator ${\mathcal{M}}(\cdot)$. In order to solve Eqs.\eqref{sec1_eq1} and \eqref{sec1_eq2}, we need to construct global surrogate models over the entire design domain $\mathcal{D}_{\boldsymbol{x}}$. \textit{Global} surrogate model differs from \textit{local} surrogate model, which ``only'' requires local information of the original simulator to solve the problem. Local surrogates can be applied to optimization problems (e.g., search the global minima \cite{jones1998efficient}) or reliability analysis (i.e., a performance-based classification problem \cite{Echard2011}). The study \cite{crombecq2011novel} shows that two essential ingredients are required to build surrogate models: (\textit{i}) a function space with certain properties and (\textit{ii}) a Design of Experiments (DoE). 
The first ingredient implies specifying a function space  $\mathcal{F}=\{\mathcal{D}_{\boldsymbol{x}} \mapsto  \mathbb{R}\}$, where, based on some loss functions, we identify an optimal function that closely resembles the original simulator. The properties of $\mathcal{F}$ rely on the selected types of surrogate models and their configurations and parameters. In the literature, one can find a wide range of surrogate models; an incomplete list includes polynomial response surfaces \cite{faravelli1989response}, artificial neural networks \cite{Papadrakakis2002}, Kriging (also called Gaussian process) \cite{Kaymaz2005}, support vector machines \cite{Basudhar2008a}, polynomial chaos expansions (PCE) \cite{blatman2010adaptive}.  The configuration for the surrogate model determines the flexibility of the function space $\mathcal{F}$. A function space with limited flexibility may not include a good approximation to the true model  $\mathcal{M}(\cdot)$. Conversely, using an excessively flexible function space increases the efforts of finding the optimal approximation and may result in overfitting (i.e., the model performs well on the training data but fails to generalize to unseen data).
%
The other ingredient, the DoE, helps the quantification of the loss functions and thus the identification of the best approximation in $\mathcal{F}(\mathcal{D}_{\boldsymbol{x}})$. A DoE consists of a set of input-output pairs, where the inputs span the design domain, and the outputs are evaluated through simulators at corresponding inputs. The core of generating a DoE is to determine the number and locations of input samples. The well-known examples of generating DoEs can be traced back to designs such as Full Factorial or Fractional Factorial \cite{kenett2021modern}, and Central Composite Design \cite{plackett1946design}. These classical DoEs were initially developed for physical experiments \cite{Garud2017}. However, DoEs for computer simulations are very different in nature as the samples are often noise-free (i.e., the outcomes come from deterministic computer simulators), and the control variables are clear and pre-defined.
It is no surprise that modern DoEs have surged dramatically in the past several decades due to the popularity of computer experiments. In this study, we investigate two categories of modern DoEs: uniform design and active-learning (a.k.a adaptive) design. 

The uniform design aims to generate samples that uniformly fill both the design space $\mathcal{D}_{\boldsymbol{x}}$ and any possible subspace of  $\mathcal{D}_{\boldsymbol{x}}$. In the literature \cite{pronzato2012design}, the uniformity in subspace is called projection or non-collapsingness property. The uniform design is achieved by optimizing some metrics that can quantify the space-filling capability of any given sample set. The study \cite{Garud2017} concludes two broad types of uniformity metrics, i.e., discrepancy-based \cite{hickernell1998generalized,caflisch1998monte} and distance-based \cite{johnson1990minimax,morris1995exploratory}. The discrepancy-based criteria measure the difference between the CDF of the uniform distribution in design space and the empirical CDF of samples, e.g., Sobols' sequence \cite{sobol1967distribution} and Halton sequence \cite{halton1964algorithm}. The distance-based criteria study the relative geometric position of samples, e.g., maximin/minimax distance design \cite{johnson1990minimax}, Voronoi Diagram-based design \cite{villagran2016non}, and minimum potential energy design \cite{eglajs1977new}. Besides, the projection property motivates researchers to study the class of so-called Latin hypercube sampling (LHS) \cite{mckay2000comparison}, which owns perfect uniformity in any one-dimensional projection. However, LHS may present poor space-filling ability in the whole design space. A variety of studies attempt to enhance the LHS by integrating the aforementioned uniformity metrics \cite{Garud2017}, e.g., maximin-LHS \cite{grosso2009finding}. In practice, the optimized size of DoE is generally unknown and it is more common to increase the size gradually until the surrogates are well-trained. This restriction necessitates a new class of quasi-uniform design that can sequentially enrich a DoE while keeping a nearly-optimized uniform metric; examples include the sequential maximin distance design \cite{stinstra2003constrained} and sequential LHS \cite{wang2003adaptive}. 

The active-learning (AL) design  selects the most informative points sequentially. It follows that AL minimizes the size of the DoE for building accurate surrogates.  The key idea of AL was initially brought from the machine learning community \cite{settles2009active}, but its earlier implementation on surrogate models can be found in Bayesian global optimizations \cite{jones1998efficient}. The method relies on the ``score functions" (e.g., the expected improvement (EI) function \cite{jones1998efficient}), which can measure the potential benefits at unexplored positions in improving the current solution. More recently, AL was introduced into reliability analyses in \cite{bichon2008efficient}, which proposed an efficient global reliability analysis (EGRA) method. In this work, the score function EI was adapted into an expected feasibility function (EFF), which can measure the deviation of the approximated model from the zero-contour of $\mathcal{M}(\boldsymbol{x})$. EGRA inspired the AL-Kriging-Monte Carlo Simulation (AK-MCS)  \cite{Echard2011}, where the MCS estimator and the adaptive surrogate construction are coupled. 
In principle, an extension of AK-MCS to full distribution estimations is straightforward since calculating CDF or CCDF at a fixed threshold $y_m$ can be considered as a standard reliability analysis problem. However, a naive implementation based on a sequence of $y_m$ significantly increases the computational costs. This computational problem was tackled in \cite{Wang2020} using a mesh-free AL-based Gaussian process (AL-GP). Specifically, the study introduced a two-step learning function to enhance efficiency. The first step identifies an optimal threshold $y^*$ by maximizing the localized full distribution function error at each AL iteration. The second step uses a reliability-based learning function (see examples in \cite{bichon2008efficient,Echard2011}) specified with the predetermined threshold $y^*$ to add a new training point into the DoE. The two-step function leverages more information about the estimated full distribution to speed up the training process of surrogate models.

Outside the UQ community, global adaptive surrogate models have also been extensively studied in the contexts of sensitivity analysis \cite{jones2008finite} and parameter estimations \cite{Mosbach2012}. Interested readers may refer to the review papers \cite{liu2018survey,fuhg2021state} for more details. These types of global surrogates are generally established on a box-constrained uniform space of interest. In UQ problems, the design space is generally not uniform but is defined by the joint PDF of input variables. In this context, the surrogates are generally built on an important region that contains the bulk of the probability mass. It follows that the parametric region of interest is typically given by an iso-probability contour rather than a box-type domain. Therefore, to conduct comparisons between AL and uniform design, we require an approach that generates uniform samples in design spaces defined by iso-probability boundaries. Observe that this is different from generating samples from the joint PDF.  

Most of the scientific literature shows the superiority of AL techniques in constructing global surrogate models compared to uniform designs \cite{farhang2005bayesian,busby2007hierarchical,li2010accumulative,crombecq2011novel,aute2013cross}. However, a few studies pointed out that AL sampling can be outperformed by the uniform design \cite{jin2002sequential}. In the context of full distribution computations, there is a research gap to answer the critical question of whether AL techniques can consistently outperform uniform design. Practically, it remains to be answered whether AL necessarily results in computational cost savings.  Furthermore, a comprehensive study to determine the influence of different components, such as the DoE enrichment and stopping criteria, on surrogate performances is missing. Therefore, the first goal of this study is to systematically compare AL and uniform design approaches in the context of full distribution computations and determine whether AL consistently outperforms uniform design.
Building on the results, the second goal of this work is to provide practical guidance for constructing global surrogate models for full distribution estimations. 

This study addresses the two goals as follows: we leverage the method presented in \cite{Wang2020} as a foundation and  then develop a modular framework, drawing upon the insights provided in \cite{moustapha2022active}. The framework is designed by combining three independent modules: a surrogate module, a DoE enrichment module, and a stopping criteria module.  The surrogate module is composed by three classical surrogate techniques: Gaussian Process \cite{sacks1989design}, Polynomial Chaos Expansion (PCE), \cite{sudret2008global} and PCE-Kriging (PCK) \cite{schobi2015polynomial}. The DoE enrichment module is composed of three learning functions and a new distance-based uniform design method, which is proposed to serve as the baseline.
Specifically, the three learning functions are: the maximum of variance (MoV) learning criterion, the two-step learning function  proposed in \cite{Wang2020}, and a gradient-based learning function proposed in \cite{mo2017taylor}\footnote{ This learning function has been developed for global surrogates but without integrating the information from the distribution function}. Moreover, the proposed uniform design method combines the maximum-minimum (maximin) distance design proposed in \cite{johnson1990minimax} with the pool-based representation as adopted in AK-MCS \cite{Echard2011}.
This sampling method is named as \textit{pool-based maximin-distance design} and it allows a sequential generation of quasi-uniform samples in non-box-constrained design spaces. 
Finally, the stopping criteria module is composed by two classes of stopping criteria: the static-based and the variance-based.  In this module, we further explore the parameter settings, such as the threshold to trigger the stopping condition and the number of consecutive triggers for the stopping condition. A combination of these selected methods is applied to solve a collection of 16 benchmark problems that involve analytical functions and FEM examples. In total, 1,920 UQ analyses are carried out to support this comparative study.

The study is organized as follows. Section \ref{sec2} presents the general framework for constructing global surrogate models and introduces the selected methods for each module. Section \ref{sec3}  develops the pool-based maximin-distance design, a new sequential uniform design method proposed as the baseline of this comparative study. Section \ref{sec4}  carries out a comparative study across a wide range of UQ problems and selected solutions. Section \ref{sec5} discusses the results  and provides recommendations on constructing surrogates for computing full distributions. Finally, Section \ref{sec6}  concludes this study.
 
\section{Methodology: sequential surrogates for computing full distributions}
\label{sec2}
\noindent
This section presents a modular framework for computing the full distribution through surrogate models. Specifically, we adapt the modular framework introduced in \cite{moustapha2022active} to our context.

\subsection{Procedure of sequential surrogate modeling}
\label{sec2_1}
\noindent
Sequential surrogate models build on an iterative approach which entails a sequence of surrogate models. The sequence starts with an initial DoE, i.e. a parsimonious sequence of input-output pairs, $\left\{\mathcal{X},\mathcal{Y}\right\} = \left\{ \left(\boldsymbol{x}_{i}^{d} ,\mathcal{M}(\boldsymbol{x}_{i}^{d}) \right) | i=1,2,..,N_{\text{d}} \right\}$.
In subsequent iterations, new surrogate models are created based on an expanded DoE. This expansion, or enrichment, is achieved by adding additional input-output pairs to the original DoE. The enrichment strategy used is designed to identify the most effective candidates for improving the training of the surrogate model. The process continues until predetermined stopping conditions are met. Ultimately, the size and position of the DoE are determined automatically, and the final updated surrogate model is used to estimate the full distribution. In summary, the steps for constructing sequential surrogate models are outlined in Algorithm \ref{table:Algotihem 1}.

\begin{algorithm}[ht]
\caption{The procedure of constructing sequential surrogates}
\label{table:Algotihem 1}
       \begin{enumerate}[left=0.5em]
       \item  \textit{Initialization}. Generate a large   candidate pool $S=\{  \boldsymbol{x}_i^{{S}}|i=1,2,..,N_{\text{MCS}} \}$ for training from the joint PDF $f_{\boldsymbol{X}}(\boldsymbol{x})$. Define an initial DoE $\left\{\mathcal{X},\mathcal{Y}\right\}$ of size $N_{\text{d}}<<N_{\text{MCS}}$.
       
       \item \textit{Training}. Train the surrogate model based on the current DoE. 
       
       \item  \textit{Compute {CDF/CCDF}}. Estimate the CDF in Eq.\eqref{sec1_eq1} and CCDF in Eq.\eqref{sec1_eq2} by Monte Carlos simulation with the surrogate model.
       
       \item \textit{Check Convergence}. If a specified stopping condition is satisfied, terminate the algorithm. 
       
       \item \textit{Enrich the DoE}. Add new training points into the DoE and proceed to Step 2. 
       
       \end{enumerate}
\end{algorithm}

\subsection{Surrogate modeling techniques}
\label{sec2_2}
\noindent
Surrogate modeling represents the core of the framework. We examine three popular surrogate models: GP, PCE, and PCK and integrate them into a unified form. For a detailed  introduction to these surrogate models, readers are referred to \cite{UQdoc_14_109,UQdoc_20_105,UQdoc_20_104}. We start with GP and later show that the other two surrogate models  share a similar formulation. 

The homogeneous GP treats the black-box simulator $\mathcal{M}(\boldsymbol{x})$ as a realization of  a Gaussian  process:
\begin{equation}
\label{eq_sec2_1}
\hat{\mathcal{M}}^{\mathrm{GP}}(\boldsymbol{x}) = \mu(\boldsymbol{x}|\boldsymbol{\beta}) +  \sigma^2 Z(\boldsymbol{x}|\boldsymbol{\theta}),
\end{equation}
where the mean/trend function $\mu(\boldsymbol{x}|\boldsymbol{\beta})$ of the Gaussian process is a regression model with  parameters $\boldsymbol{\beta} \in \mathbb{R}^q$, $\sigma^2$ is the process variance, and $Z(\boldsymbol{x}|\boldsymbol{\theta})$ is a zero-mean, unit-variance Gaussian process uniquely determined by a homogeneous correlation function/kernel $R(\boldsymbol{x}_{i},\boldsymbol{x}_{j}|\boldsymbol{\theta}) \in\mathbb{R}$ parameterized by $\boldsymbol{\theta}$.  Commonly used kernels include the Gaussian, cubic, spline, and Matérn (see \cite{lophaven2002dace} for details).  It is typical to assume a linear regression model for $\mu(\boldsymbol{x}|\boldsymbol{\beta})$, that is $\mu({\boldsymbol{x}}) = \boldsymbol{\Psi}^{\mathrm{T}}(\boldsymbol{x})\boldsymbol{\beta}$,  where the vectors $\boldsymbol{\Psi}(\boldsymbol{x}) \in \mathbb{R}^{q}$  contains $q$ predefined basis functions.


Given the prior Gaussian assumption in Eq.\eqref{eq_sec2_1}, the GP model parameters $\boldsymbol{\beta}$ and $\boldsymbol{\theta}$ are tuned to optimally fit the observed data. Here, the optimal fit requires the GP to be a Best Linear Unbiased Prediction (BLUP) \cite{santner2003design}, i.e., the prediction needs to be linear ($\hat{Y}_\mathrm{L}(\boldsymbol{x}) = \boldsymbol{c}^\mathrm{T}\mathcal{Y}$ with $\boldsymbol{\boldsymbol{c}}=\boldsymbol{c}(\boldsymbol{x})\in \mathbb{R}^{N_d}$ and $\mathcal{Y}\in \mathbb{R}^{N_d}$ is the observed output vector), unbiased (\(\mathbb{E}[\hat{Y}_L(\boldsymbol{x})-\hat{\mathcal{M}}^{\mathrm{GP}}(\boldsymbol{x})] = 0\)), and best ($\underset{\boldsymbol{c}}{\min}\  \mathrm{MSE}(\boldsymbol{x}) = \mathbb{E} [(\hat{Y}_L(\boldsymbol{x})-\hat{\mathcal{M}}^{\mathrm{GP}}(\boldsymbol{x}))^2]$).\footnote{In our study, the GP predictor is interchangeable with the Kriging predictor as defined in \cite{santner2003design,lophaven2002dace}. Some studies (e.g., \cite{Wang2020} ) refer to the GM predictor as the mean of the posterior distribution of Eq.\eqref{eq_sec2_1} conditional on observations ${\mathcal{X}}$ and $\mathcal{Y}$. While both methods offer the same prediction to unexplored points,  their prediction errors/uncertainties differ.} Given these constraints, the GP predictor has the analytical form \cite{santner2003design}:
\begin{equation}
\label{eq_sec2_2}
\begin{aligned}
    \hat{\mathcal{M}}({\boldsymbol{x}})  & = \boldsymbol{\Psi}^{\mathrm{T}} (\boldsymbol{x})\boldsymbol{\hat{\beta}} + \mathbf{r}^\mathrm{T}(\boldsymbol{x}|\boldsymbol{\theta}) \mathbf{R}^{-1} (\mathcal{Y}-\mathbf{F}\boldsymbol{\hat{\beta}}) \\
    & = \boldsymbol{\Psi}^{\mathrm{T}} (\boldsymbol{x})\boldsymbol{\hat{\beta}} + \mathbf{r}^\mathrm{T}(\boldsymbol{x}|\boldsymbol{\theta}) \boldsymbol{\hat{\gamma}}\,,
\end{aligned}
\end{equation}
where $\boldsymbol{\hat{\beta}}=\boldsymbol{\hat{\beta}}(\boldsymbol{\theta})$  is the generalized least-squares estimation of  $\boldsymbol{\beta}$, and other symbols are defined as: $[\mathbf{r}(\boldsymbol{x|\boldsymbol{\theta}})]_{i,1} = R(\boldsymbol{x},\boldsymbol{x}_{i}^{d}|\boldsymbol{\theta})$, $[\mathbf{R}]_{i,j} = R(\boldsymbol{x}_{i}^{d},\boldsymbol{x}_{j}^{d}|\boldsymbol{\theta})$ and $[\mathbf{F}]_{i,:} = \boldsymbol{\Psi}^{\mathrm{T}}(\boldsymbol{x}_i^d)$. The coefficients  $\boldsymbol{\hat\gamma} \in \mathbb{R}^{N_d}$   are constrained by requiring exact interpolations at each design point, i.e., $ \mathbf{R}\boldsymbol{\hat\gamma} =  \boldsymbol{\epsilon} $, where $\boldsymbol{\epsilon} = \mathcal{Y}-\mathbf{F}\boldsymbol{\hat{\beta}}$  are the regression residuals. The unknown hyper-parameters $\boldsymbol{\theta}$  can be calibrated by maximum likelihood estimation or cross-validation \cite{santner2003design}.

In Eq.\eqref{eq_sec2_2}, the GP predictor is decomposed as a global predictor $\hat{\mathcal{M}}^{\mathrm{G}}(\boldsymbol{x}) = \boldsymbol{\Psi}^{\mathrm{T}} (\boldsymbol{x})\boldsymbol{\hat{\beta}}$ and a local predictor $\hat{\mathcal{M}}^{\mathrm{L}}(\boldsymbol{x}) = \mathbf{r}^\mathrm{T}(\boldsymbol{x}|\boldsymbol{\theta}) \boldsymbol{\hat{\gamma}}$. Both individual predictors are expressed as a dot product between a basis function vector ($\boldsymbol{\Psi} (\boldsymbol{x})$ or $\mathbf{r}(\boldsymbol{x}|\boldsymbol{\theta})$) and a coefficient vector ($\boldsymbol{\hat{\beta}}$ or $\boldsymbol{\hat{\gamma}}$). Due to the flexibility of basis functions $\mathbf{r}(\boldsymbol{x}|\boldsymbol{\theta})$, a constant trend is assigned to the GP in this study; thus GP degenerates to a local predictor.

Classical PCE is a global predictor by setting the process variance $\sigma^2$ in Eq.\eqref{eq_sec2_1} to zero and eliminating the local terms in Eq.\eqref{eq_sec2_2}. The basis functions $\boldsymbol{\Psi}(\boldsymbol{x})$ of PCE are chosen to be multivariate orthogonal polynomials with respect to the input PDF.  Once a set of basis functions is determined, the PCE coefficients $\boldsymbol{\beta}$ can be computed using algorithms such as the projection method, least-square regression, or least angle regression \cite{UQdoc_20_104}. 

PCK (PCE-Kriging) is interpreted as a GP/Kriging model with PCE as the trend function $\mu(\boldsymbol{x}|\boldsymbol{\beta})$ in Eq.\eqref{eq_sec2_1}.  The construction of PCK involves two steps: (i) select an optimal set of multivariate polynomials for the trend functions, and (ii) use the standard calibration procedure of GP to train PCK. As a result,  PCK shares the same formulation as GP in Eq.\eqref{eq_sec2_2} \cite{schobi2015polynomial}.

In this study,  both GP and PCK provide the built-in error/prediction uncertainty measure $\sigma^2_{\hat{\mathcal{M}}}(\boldsymbol{x})$ due to the Gaussian process assumption. PCE requires other techniques to estimate the surrogate errors, such as bootstrap resampling strategy \cite{Marelli2018}, $k$  fold cross-validation \cite{Xiao2018}, semi-variance \cite{seung1992query}, gradient-based variance \cite{crombecq2011novel}. This study uses the bootstrap resampling method proposed in \cite{Marelli2018} to estimate the local error  $\sigma^2_{\hat{\mathcal{M}}}(\boldsymbol{x})$ of PCE.

\subsection{Error measure}
\label{sec2_3}
\noindent
We adopt an error measure proposed in \cite{Wang2020} for measuring the discrepancy distance between two CDFs, denoted as $F_{Y}(y)$  and $\hat{F}_{Y}(y)$. This metric owns a symmetry property in evaluating CDFs and CCDFs simultaneously. In addition, it emphasizes the contribution from the tails of the full distribution. The measure is formulated as
\begin{equation}
\label{eq_sec2_3}
{\epsilon}_{F}(F_{Y}(y),\hat{F}_{Y}(y))=\frac{1}{y_{\max}-y_{\min}} \times \int_{y_{\min}}^{y_{\max}} w(y)dy,
\end{equation}
and
\begin{equation}
\label{eq_sec2_4}
w(y) = {\frac{\left| {F}_{Y}(y)-\hat{F}_{Y}(y) \right|}{\min \left[ {F}_{Y}(y),1-{F}_{Y}(y) \right]}},
\end{equation}
where the integrand $w(y)$ measures the absolute relative distance between  $F_{Y}(y)$  and $\hat{F}_{Y}(y)$ at position $y$. The  integral in Eq.\eqref{eq_sec2_3} accounts for the accumulated errors on the integral range $[{{y}_{\min }},{{y}_{\max }}]$. The error measure is finally defined as the integration of $w(y)$  divided by $(y_{\max}-y_{\min})$, i.e., the average of relative absolute errors over the range of interest. This error measure is used to assess the accuracy of surrogates and construct stopping criteria in the next subsection.

\subsection{Stopping criteria}
\label{sec2_4}
\noindent
The stopping criteria are used to terminate the iterative training of sequential surrogates. Conservative stopping criteria could result in redundant DoE points, while relaxed criteria lead to inaccurate surrogates. Similar to \cite{moustapha2022active}, this work investigates two classes of stopping criteria related to the convergence trend of the error measure in Eq.\eqref{eq_sec2_3}. 

The first type is named the static-based criterion. This criterion measures the stability of the estimated CDF in successive iterations and it is written as:
\begin{equation}
\label{eq_sec2_5}
{\epsilon}_{S} = {\epsilon}_{F}(\hat{F}^{(i-1)}_{Y}(y),\hat{F}^{(i)}_{Y}(y)) \leq \epsilon^{tol}_{S},
\end{equation}
where the  $F^{(i)}_{Y}(y)$ is the estimated CDF at $i$th iteration of  sequential surrogates. This iteration stops  when the condition in Eq.\eqref{eq_sec2_5} is satisfied. 

The second type is named as variance-based criteria which was originally investigated in \cite{Wang2020}. This criterion stops the training when the confidence bound of the estimated full distribution is sufficiently narrow. This criterion requires an estimate of the prediction error and is defined as:
\begin{equation}
\label{eq_sec2_6}
{\epsilon}_{V}(\hat{F}^+_{Y}(y),\hat{F}^0_{Y}(y),\hat{F}^-_{Y}(y))=\frac{1}{y_{\max}-y_{\min}} \times \int_{y_{\min}}^{y_{\max}}  {\frac{\left| \hat{F}^+_{Y}(y)-\hat{F}^-_{Y}(y) \right|}{\min \left[ \hat{F}^0_{Y}(y),1-\hat{F}^0_{Y}(y) \right]}}dy \leq \epsilon^{tol}_{V},
\end{equation}
where $\hat{F}^+_{Y}(y)$, $\hat{F}^0_{Y}(y)$ and $\hat{F}^-_{Y}(y)$  are the CDFs  calculated by the bounded surrogates $\hat{\mathcal{M}}^k(x)=\hat{\mathcal{M}}(x)+k\sigma_{\hat{\mathcal{M}}}(\boldsymbol{x})$ with $k$ setting as 2, 0, -2, respectively. Eq.\eqref{eq_sec2_6} is still developed based on the error measure in Eq.\eqref{eq_sec2_3}. 

For both stopping criteria, we can enhance their robustness by requiring them to be consecutively met for multiple iterations, such as 2 or 3 times. In summary, the thresholds $\epsilon^{tol}_{S}$ in Eq.\eqref{eq_sec2_5}, $\epsilon^{tol}_{V}$ in Eq.\eqref{eq_sec2_6}, and the required number of consecutive met criteria are hyper-parameters in this study. Section \ref{sec4_4} discusses the tuning of these parameters.

\subsection{Active learning for DoE enrichment}
\label{sec2_5}
\noindent
AL strategies for DoE enrichment are based on an iterative process. Specifically, a single sample is chosen at each step to maximize the learning function. The surrogate model is then retrained, and the training set is updated. This process is repeated until specific stopping criteria are met. The learning function completely defines the active learning strategy.

This subsection introduces three learning functions used in this paper. For simplicity, here, we report a brief yet self-contained description. More details are provided in the supplementary materials (see  \ref{Appedndix_A}). 
The three learning functions are:

\begin{itemize}
\item Maximum of Variance (MoV) learning function. 
\item Two-step learning function \cite{Wang2020}.
\item Gradient-based learning function \cite{crombecq2011novel,mo2017taylor}.
\end{itemize}
The MoV learning function relies on the maximum variance criterion. Specifically, the best point is the one with the highest variance, $\sigma^2_{\hat{\mathcal{M}}}(\boldsymbol{x})$. In GP- and PCK-based surrogate models, the variance is explicitly known, whereas, in classical PCE, it is estimated through bootstrap resampling techniques.

The two-step learning function proposed in \cite{Wang2020} includes two steps to select the optimal samples. In the first step, a threshold $y^*$ is selected, which contributes the most errors/uncertainties to the full distribution estimation.  In the second step, a candidate point is selected using a given learning function (e.g., a reliability-based AL function \cite{Echard2011,bichon2008efficient}). Notably, this strategy is ``mesh-free'' as it does not rely on a fixed discretization of the distribution function.

 The gradient-based learning function considers a trade-off between global exploration and local exploitation.  Global exploration involves investigating a wide area to discover strongly nonlinear regions. Local exploitation focuses on these specific regions to select the best possible samples. Specifically, the global exploration searches the regions where the DoE in design space is sparse. The local search strategy leverages gradient information to exploit locations characterized by pronounced nonlinearity (which are inherently more unpredictable). To achieve a balance  between exploration and exploitation, a weighting function is introduced. The gradient-based learning function was first studied in \cite{crombecq2011novel}. In this work, we adopt a variation introduced in \cite{mo2017taylor}, where  the gradient of the surrogate, rather than that of the original simulator, is used.

\section{Uniform design based on maximin distance}
\label{sec3}
\noindent
This section develops a uniform experimental design method, termed \textit{pool-based maximin distance design},  for constructing global surrogates. First, the existing maximin distance design method is presented. Then, this technique is adapted by incorporating the pool-based representation approach. Finally, the modified method is compared with two classical uniform design methods. 

\subsection{Original maximin-distance design}
\label{sec3_1}
\noindent
Maximum-minimum (maximin) distance design was originally proposed to uniformly allocate samples in the unit hypercube space \cite{johnson1990minimax}. Consider a design space $\mathcal{D_{\boldsymbol{x}}}\subset{{\mathbb{R}}^{N}}$ and a set of design input $\mathcal{X}=\{ \boldsymbol{x}_{i}^{d}\in\mathcal{D}_{\boldsymbol{x}}| i=1,2,...,N_d\}$. ${{\mathcal{X}}^{*}}$ is the maximin-distance design if
\begin{equation}
\label{eq_sec3_1}
\mathcal{X}^{*} =  \mathop{\arg\max}_{\mathcal{X}}\left( \,\underset{ {j}\neq{i} }{\mathrm{\min }}\,d({\boldsymbol{x}_{i}^{d}},{\boldsymbol{x}_{j}^{d}}) \right) ,
\end{equation}
where $d(\cdot,\cdot)$ is the Euclidean distance measure. Eq.\eqref{eq_sec3_1}, first, implements the ``${\min }$" operator to calculate the minimum distance among distinct points in $\mathcal{X}$. Then, the ``$\mathop{\arg\max}$" operator searches for a design ${{\mathcal{X}}^{*}}$ that maximizes this minimum distance.  Here, the optimality implies that the distances from each point  $\boldsymbol{x}^d_i  \subset\mathcal{X}$ to its nearest point are equal, thus the design space is uniformly filled.

The objective function in Eq.\eqref{eq_sec3_1} is highly nonlinear and non-convex with $N_d\times N$ variables \cite{stinstra2003constrained}. When $N_d\times N$ is large, the optimization problem is computationally unfeasible. Moreover, extending the maximin-distance design to design spaces with non-box and no-compact sets is challenging. A feasible solution is given by solving the optimization problem sequentially \cite{stinstra2003constrained}. The idea lies in decomposing the original problem into sub-optimization problems and solving them sequentially. Each sub-optimization problem focuses on optimizing a small number of points based on the maximin-distance principle.  Next, the optimized points from previous sub-problems are fixed for the next sub-problems. In a special case, we can optimize a single point in each subproblem. This strategy is straightforward, and the next subsection introduces the combination with the pool-based representation to further simplify the optimization process.

\subsection{Pool-based maximin-distance design}
\label{sec3_2}
\noindent
To build surrogate models, the method AK-MCS \cite{Echard2011} utilized a combination of the AL and the candidate-pool-based representation. The candidate pool can represent the design space by using MCS to simulate a set of candidate points, among which AL techniques are applied to identify the optimal points. 
This study introduces a sequential maximin-distance design approach applied to an MCS pool, motivated by the following ideas:
\begin{itemize}
    \item \textbf{Computational efficiency:} Representing the design space as a pool of candidate samples transforms the continuous optimization problem (Eq.\eqref{eq_sec3_1}) of optimizing positions  into a discrete optimization problem of selecting $N_d$ points in the pool, leading to significant computational savings.
    \item \textbf{Integration with input PDFs:} The pool-based method implicitly accounts for the distributions of the input variables, enabling the surrogate model training to concentrate on regions with non-negligible probabilities.
    \item \textbf{Improved projection properties:} While discretization and sequential sampling may reduce the uniformity of the input DoE, they enhance the uniformity of projection onto subspaces when compared to the original maximin-distance design, as documented in \cite{joseph2016space}.
    \item \textbf{Robustness to design space boundaries:} Representing the design space by samples makes the optimization process insensitive to the complexity of the design space boundaries.

    \item \textbf{Compatibility with active learning:} The pool-based method allows for a sequential enrichment of the DoE, where the optimal size of the DoE is not predetermined. 
\end{itemize}

The proposed \textit{pool-based maximin-distance design} is reported in Algorithm \ref{table:algorithm 2}.

\renewcommand{\algorithmcfname}{Algorithm}
\begin{algorithm}[ht]
\caption{Procedure of pool-based maximin-distance design}
\label{table:algorithm 2}
\begin{algorithmic}
\STATE
       \begin{enumerate}[left=0.5em]
       \item \textit{Generate a candidate pool $S=\{  \boldsymbol{x}_i^{{S}}|i=1,2,..,N_{\rm{MCS}} \}$}.
       \item \textit{Normalize the candidate pool}. Apply component-wise Z-score normalization to each sample in  $S$, obtaining  $\tilde{S}=\{  \boldsymbol{\tilde{x}}_i^{{S}}|i=1,2,..,N_{\text{MCS}}\}$.
       \item \textit{Build an initial DoE}. Choose a random point from $\tilde{{S}}$ to obtain the current DoE  $\tilde{\mathcal{X}} \leftarrow \{{\tilde{{x}}}^d_{1}\}$ and set $N_d\leftarrow1$.
       \item \textit{Enrich the DoE by sequential maximin-distance design}. For each iteration, chooses a point in $\tilde{S}$ with the maximum distance to $\tilde{\mathcal{X}}$, i.e.,  
\begin{equation*}
\begin{gathered}
\label{eq_sec3_2_1}
   \tilde{x}^*=\mathop{\arg\max}_{{{{\boldsymbol{x}}_{i}}\in\tilde{S}}}\left(\min_{\boldsymbol{x}_j\in\tilde{\mathcal{X}}}d({{\boldsymbol{x}}_{i}},{{\boldsymbol{x}}_{j}})\right)\,.
\end{gathered}
  \end{equation*}
  Then, $\tilde{\mathcal{X}}\leftarrow\tilde{\mathcal{X}}\cup\{\tilde{x}^*\}$, $N_d\leftarrow N_d+1$. Repeat this enrichment process until $N_d$  reaches a prescribed number. 
       \item\textit{Reverse the normalization}. Perform inverse Z-score normalization to  obtain $\mathcal{X}$.
       \end{enumerate}
\end{algorithmic}
\end{algorithm}

Algorithm \ref{table:algorithm 2} entails three important aspects: 

\begin{itemize}
\item \textbf{Parallel processing:} The maximin-distance design can add multiple points to a DoE without evaluating the simulators  $\mathcal{M}(\cdot)$. This feature can be especially advantageous when parallel computing resources are available to construct surrogates.
\item \textbf{Markov Chain Monte Carlo-based candidate pool generation:} The candidate pool can be generated using Markov Chain Monte Carlo techniques, when the input distribution contains an unknown normalizing constant. This is relevant in Bayesian problems.

\item \textbf{Deterministic design:} If the candidate pool in Step 1 and the initial point in Step 3 are fixed, this method transforms into a deterministic design, ensuring a consistent DoE that can be used to compare the performance of different surrogate models.


\end{itemize}


\subsection{Comparative study of uniform designs}
\noindent
This subsection compares the pool-based maximin-distance design (referred to as MD hereafter) with two other uniform designs: Latin Hypercube Sampling (LHS)  \cite{mckay2000comparison} and Sobol sequence design (Sobol)  \cite{sobol1967distribution}. The comparison involves visually inspecting their uniformity in two 2-dimensional design spaces: a unit cube space and a bi-variate standard normal space. Additionally, we compare their performance in assisting surrogate modeling for full distribution estimation. For MD, a pool size of $10^5$ is used. 

\begin{figure}[ht]
   	\begin{center}
	\includegraphics[width=16cm]{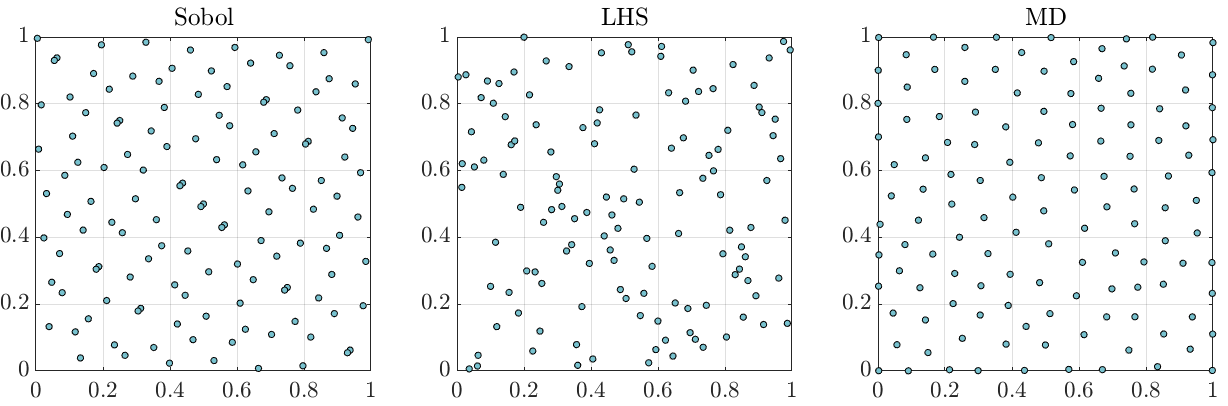}
	\caption{DoEs in the unit cube using three uniform design methods. \textit{The sample size is 128}.}
	\label{fig1}
	\end{center}
\end{figure}

Firstly, we compare the samples generated by the three uniform design methods in the 2-dimensional unit cube. The results are shown in  Fig.\ref{fig1}. Sobol sequence demonstrates good global uniformity but exhibits some local clustering along the diagonals. The samples of LHS appear fairly random and lack uniformity. Meanwhile, MD outperforms Sobol sequence and LHS in terms of uniformity and allocates more points along the boundaries.

\begin{figure}[ht]
   	\begin{center}
	\includegraphics[width=16cm]{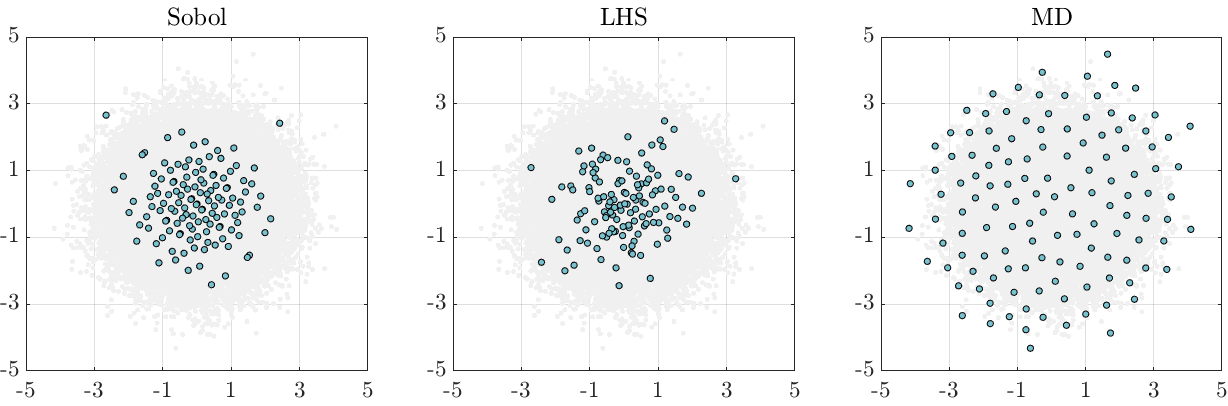}
	\caption{DoEs in the standard normal space using three uniform design methods. \textit{The sample size is 128.}}
	\label{fig2}
	\end{center}
\end{figure}

Next, a comparison is conducted in the standard normal space. In this case, Sobol sequence and LHS generate samples through  iso-probability transformations from the unit cube space to the standard normal space. For MD, the samples are selected from a pool of $10^5$ random bi-variate standard normal samples. The DoEs and the pool of $10^5$ samples are shown in Fig.\ref{fig2}. It is observed that samples from the Sobol sequence and LHS tend to concentrate around the mode. This is expected because they are quasi-Monte Carlo techniques designed to mimic the distribution. In contrast, MD sampling is designed for training surrogate models, thereby distributing points more uniformly. This feature is crucial for the construction of global surrogates to explore both the mode and the tail regions.

\begin{figure}[ht]
   	\begin{center}
	\includegraphics[width=12cm]{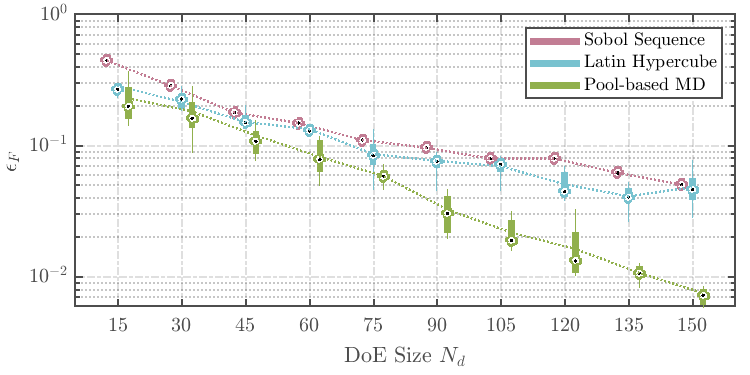}
	\caption{A comparison of surrogate modeling accuracy for distribution estimation using different uniform designs. {\textit{The error measure $\epsilon_F$ is described by Eq.\eqref{eq_sec2_3}. }}}
	\label{fig3}
	\end{center}
\end{figure}

Finally, we compare the three uniform design methods in terms of surrogate modeling accuracy using an expanding DoE, as illustrated in Fig.\ref{fig3}. This comparison uses a 6-dimensional analytical model with multiple types of input distributions: the normal, Weibull, and uniform. Details of this computational model are provided in benchmark example \#11, as referenced in \ref{Appedndix_A}. Since MD and LHS are stochastic methods, the prediction variability is evaluated based on ten independent runs. The surrogate model is a Gaussian process with a constant trend and a Matérn-5/2 kernel. The figure indicates that MD leads to highly accurate and efficient surrogate modeling. Given this superior performance in building surrogates, we leverage MD as a baseline to assess the potential improvements of active learning over uniform designs.

\section{Comparative study}
\label{sec4}
\subsection{Preliminary work}
\label{sec4_1}
\subsubsection{Benchmark problems set-up}
\label{sec4_1_1}
\noindent
The benchmarks consist of 16 examples drawn from the literature. Most of the examples are formulated in analytical forms, with the exception of three that employ FEM models. The FEM models include a 21-bar truss structure \cite{blatman2010efficient}, a 3-bays-5-floor frame structure \cite{liu1991kiureghian}, and a 10-story shear structure \cite{ohtori2004benchmark}. The number of random variables varies from 2 to 21, covering the low- and medium-dimensional problems.  A detailed list of the benchmarks, including the performance functions, probabilistic models, and the ranges of QoIs $[y_{\min},y_{\max}]$ are summarized in the supplementary material (see \ref{Appedndix_A}). In this study, the range $[y_{\min},y_{\max}]$ is defined as $[F^{-1}_Y(10^{-k}), F^{-1}_Y(1-10^{-k})]$, where  $F^{-1}_Y(\cdot)$ is the inverse function of the target CDF, and we set $k=3$. This range contains $99.98\%$ of the  probability mass, ensuring accurate estimation of lower statistical moments and satisfying the needs of most engineering applications. Notice that these bounds are computed using an empirical CDF obtained from MCS. However, in real-life examples, the target distribution is not available. In this case, \cite{Wang2020} provides algorithmic solutions to estimate $[y_{\min},y_{\max}]$.

\subsubsection{ Algorithmic settings}
\label{sec4_1_2}
\noindent
This subsection describes in detail the parameters  and  configurations settings related to Algorithm   \ref{table:Algotihem 1}.  In \textit{Step 1}, the initial DoE is generated by the pool-based MD. This choice is motivated by  \cite{ma2022novel}, which shows that space-filling sampling for initial DoEs is  more robust than random sampling methods. In particular, an MCS population of size  $10^{k+2}$ is drawn, from which an initial DoE is extracted using the pool-based MD. The MCS population is later used for computing statistics of interest, with the parameter $k+2$  chosen to ensure the coefficient of variation of tail estimations at approximately 10\%.
The size of the initial DoE is set as $\max\{12,3N\}$, where $N$ denotes the dimension  of the problem. We set 3$N$  so that there are at least 3 points for each direction to capture possible nonlinearities\footnote{If the computational cost is too high, one might opt for 2$N$}.  In\textit{ Step 2}, the settings of the surrogate models, GP, PCE, and PCK,  are specified based on \cite{moustapha2022active}.  
In \textit{Step 3},  the range of interest $[y_{\min},y_{\max}]$ is equally divided into 100 intervals to compute the full distribution as outlined in Eqs.\eqref{sec1_eq1} and \eqref{sec1_eq2}. 
In \textit{Step 4}, the stopping criteria module consists of two ways to terminate the iterative process: the first is introduced in Section \ref{sec2_4}, and the second  sets the maximum DoE size to $\min\{100+N\times 20,300 \}$. The first criterion entails hyper-parameters such as thresholds and the number of consecutive triggers for the stopping criterion. Hence, we employ only the second criterion to stop the algorithm, facilitating the collection of data to analyze the optimal hyper-parameter settings for the first stop criterion. Additionally, we have increased the maximum model evaluations for Examples \#3, \#5, and \#6 to 300 to accommodate their slower convergence rates.
In \textit{Step 5}, the DoE enrichment module contains three AL methods introduced in Section \ref{sec2_5}, i.e., the maximum of variance (MoV), the two-step learning function (t-LF), and the gradient-based learning function (g-LF), and the uniform design method introduced in Section \ref{sec3}. Notably, these enrichment methods do not require any hyper-parameter tuning. 

To sum up, the 16 benchmark examples are solved by 12 aggregated strategies, which are designed by integrating 3  surrogates with 4 DoE enrichment methods. Each of the 12  strategies is independently run 10 times to solve the benchmarks, resulting in a total of $1,920$ UQ analyses to support the comparative study.

\subsubsection{Criteria for ranking the strategies}
\label{sec4_1_3}
\noindent
In order to identify the ``optimal''  configurations in each module, this study carries out rankings based on the statistical analysis over the $1,920$ UQ analyses. The comparative indicators include two metrics regarding the accuracy and efficiency/cost. The accuracy metric relies on the error measure $\epsilon_{F}$ in Eq.\eqref{eq_sec2_3}, and the efficiency/cost metric is formulated based on the number of model evaluations. The adopted comparative indicators vary with different modules. The \textit{surrogate module} and \textit{DoE enrichment module} are evaluated based on the \textit{mean} error measure that averages the results of the 10 repeated runs. The metrics for these two modules include: 
\begin{itemize}
\item \textit{Accuracy metric}:  Given fixed iterations (i.e., the maximum size of the DoE prescribed in Section \ref{sec4_1_2}), compute the percentage of the cases that the average error measure is smaller than a prescribed tolerance  $\epsilon^{\text{tol}}_{F}$.
\item \textit{Efficiency metric}: Given an error tolerance $\epsilon^{\text{tol}}_{F}$, compute the percentage of the cases that the selected methods use the minimum average iterations to converge (i.e., reach $\epsilon_{F} \le \epsilon^{\text{tol}}_{F}$).
\end{itemize}

The \textit{Stopping criteria module} is used to terminate the iterative process. Each stopping criterion is configured to halt the surrogate training at a specific iteration, returning the required number of model evaluations $N_{\mathcal{M}}$ and the corresponding error measure $\epsilon_{F}$. The optimal stopping criterion is the one that stops the iteration while satisfying the accuracy demand, i.e., $\epsilon_{F} \le \epsilon^{\text{tol}}_{F}$. In other words, the surrogate training achieves convergence using the minimum number of model evaluations $N^{\min}_{\mathcal{M}}$. The comparative indicators for this module are defined as follows:
\begin{itemize}
\item \textit{Accuracy metric}: Given an error tolerance  $\epsilon^{\text{tol}}_{F}$, compute the percentage of the cases (over the $1,920$ analyses) that accuracy demand is met, i.e., $\epsilon_{F} \le \epsilon^{\text{tol}}_{F}$.
\item  \textit{Cost metric}: Given an error tolerance  $\epsilon^{\text{tol}}_{F}$, the cost is defined as the mean value (over the $1,920$ analyses)  of the normalized number of model evaluations, i.e., $N_{\mathcal{M}}/N^{\min}_{\mathcal{M}}$.
\end{itemize}

\subsection{Compare the performance across surrogate models}
\label{sec4_2}
\noindent
In this subsection, we compare the performance of the surrogates prescribed with different target accuracy tolerances $\epsilon_{F}^{\text{tol}}$, which accounts for different accuracy demands in practical applications. The set $\epsilon_{F}^{\text{tol}}\in\{0.05,0.1,0.15,0.20,0.25\}$ is considered, while 0.05 corresponds to an accurate model and 0.25 a rough estimation. 

As a reminder, this work carries out a benchmark study consisting of 16 examples solved by 12 different solution strategies. These strategies are designed by combining the two independent components---3 surrogate models and 4 DoE enrichment methods. The performance of each surrogate is assessed through the 64 cases (16 examples $\times$ 4 DoE enrichment methods).

\begin{figure}[ht]
   	\begin{center}
	\includegraphics[width=10cm]{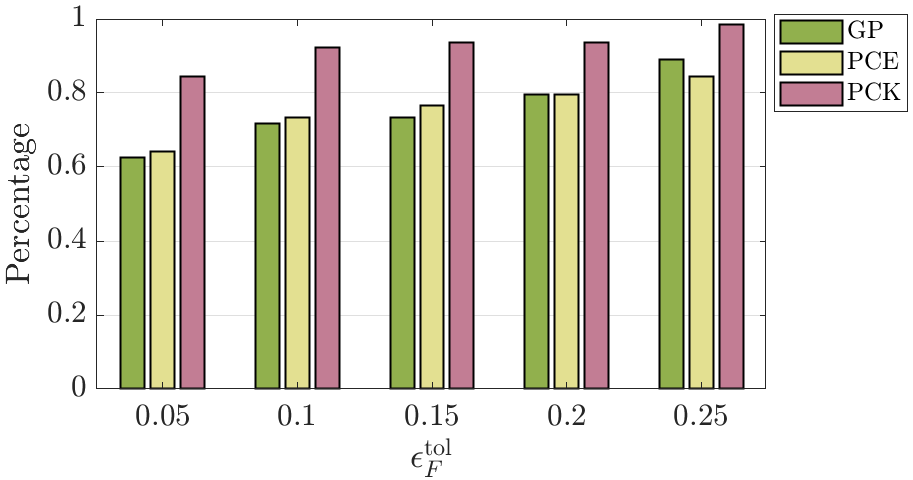}
	\caption{Comparison of surrogate models in terms of accuracy. \textit{ The accuracy metric uses the percentage of  cases (out of the 64 cases) that the surrogate model can converge (i.e., $\epsilon_{F} \le \epsilon^{\text{tol}}_{F}$) within a fixed maximum number of iterations}.}
	\label{fig5_1}
	\end{center}
\end{figure}

First, the accuracy metric is applied to rank the three surrogates. This metric starts with counting the number of occurrences of  $\epsilon_{F} \le \epsilon^{\text{tol}}_{F}$ among the 64 cases. This procedure is repeated by changing the thresholds from 0.05 to 0.25. Then the percentage of cases (out of the 64 cases) that the surrogates can converge is summarized. The results are shown in Fig.\ref{fig5_1}. We observe that, for all surrogates, the accuracy metric increases with the increase of  $\epsilon^{\text{tol}}_{F}$. This phenomenon is expected since constructing surrogates with lower accuracy is easier. In Fig.\ref{fig5_1},  surrogate  PCK consistently shows the highest accuracy, which converges in $84\%$  of the tested cases when  $\epsilon_{F}^{\text{tol}}=0.05$ and this ratio increases to $98\%$ when  $\epsilon_{F}^{\text{tol}}=0,25$. GP is slightly less effective than the other two surrogates when $\epsilon_{F}^{\text{tol}}<0.2$, and its relative performance improves when $\epsilon_{F}^{\text{tol}}=0.25$.

\begin{figure}[ht]
   	\begin{center}
	\includegraphics[width=10cm]{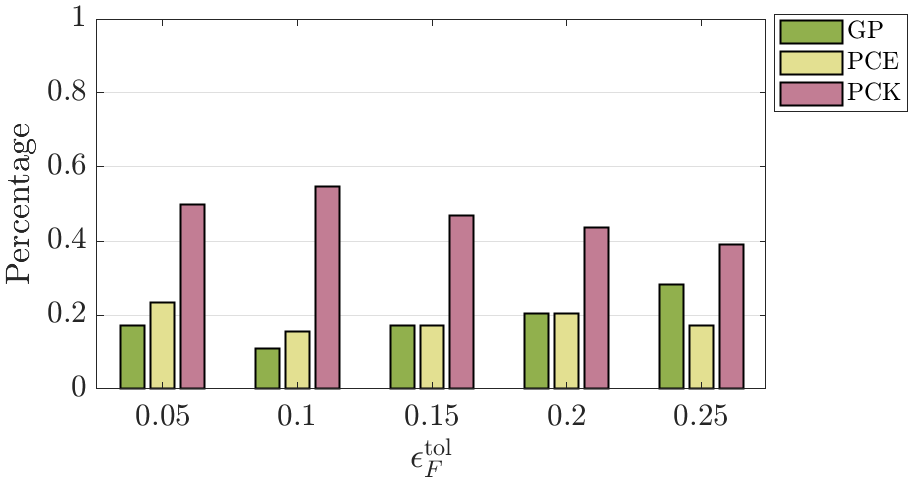}
	\caption{Comparison of surrogate models in terms of efficiency. \textit{The efficiency metric uses the percentage of  cases (out of the 64 cases) that the surrogate model requires the minimum iterations to converge (i.e., $\epsilon_{F} \le \epsilon^{\text{tol}}_{F}$) .}}
	\label{fig5_2}
	\end{center}
\end{figure}

Next, the surrogates are compared across the 64 cases regarding the efficiency metric. The surrogate using the least number of model evaluations to converge (i.e.,$\epsilon_{F} \le \epsilon^{\text{tol}}_{F}$) is labeled as the winner. Then, we compute the proportion of cases in which the given surrogate wins the comparisons. The results are presented in Fig.\ref{fig5_2}. PCK is again the best choice in terms of efficiency with different $\epsilon_{F}^{\text{tol}}$. GP outperforms PCE when $\epsilon_{F}^{\text{tol}}=0.25$, but it becomes comparable to PCE when  $\epsilon_{F}^{\text{tol}}$  is set as $0.15$ or $0.25$. Besides, GP performs less  accurately than PCE when  $\epsilon_{F}^{\text{tol}}$  goes down to 0.1 or 0.05.

\begin{figure}[ht]
   	\begin{center}
	\includegraphics[width=16cm]{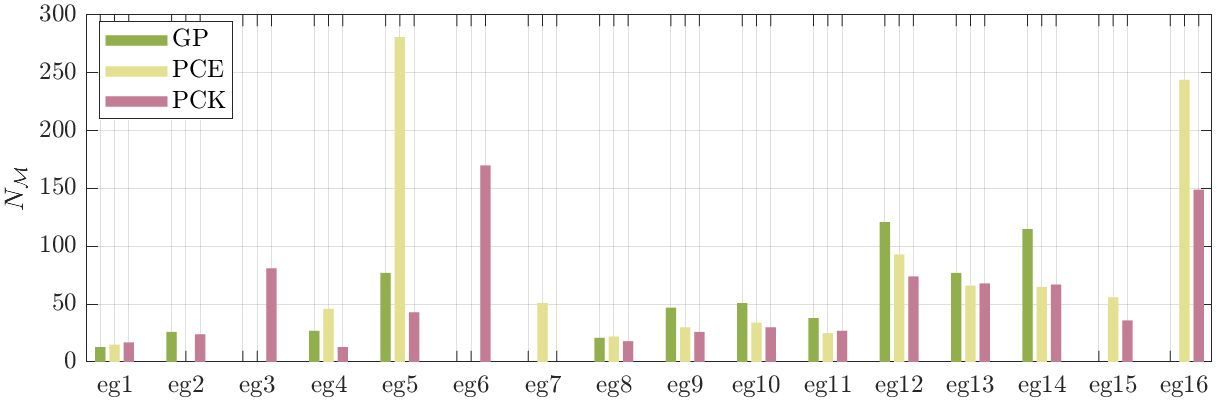}
	\caption{The average number of model evaluations required to converge ($\epsilon_{F} \le \epsilon^{\text{tol}}_{F}=0.1$). \textit{The bar is not shown if the method fails to converge within the given maximum number of iterations.}}
	\label{fig7_2}
	\end{center}
\end{figure}

From the above comparisons, we observe that for most cases PCK outperforms the other two in terms of accuracy and efficiency. To reveal more details regarding the comparison, we decompose the aggregated results in Fig.\ref{fig5_2} (with $\epsilon_{F}^{tol}=0.1$) into detailed comparisons as shown in Fig.\ref{fig7_2}, where it reports the average number of model evaluations to converge among the 16  examples. Fig.\ref{fig7_2} only shows the results from surrogates constructed by the uniform design MD. It is observed from Fig.\ref{fig7_2} that the best surrogate model varies with the problem. For example, GP is the most efficient option for Examples \#1, PCE outperforms the others in Examples \#7, \#11, \#13, \#14, and PCK shows the best performance in the remaining examples. It is also observed that the optimal surrogate model sometimes only brings marginal improvement over the alternatives, such as in Examples \#1, \#8 $\sim$ \#11, and \#13, while sometimes it is significantly better,  such as in Examples \#2, \#3, \#5 $\sim$ \#7, \#15, and \#16. In general, without prior knowledge about the problems, PCK would be the most robust choice.

\subsection{Compare the performance across DoE enrichment methods}
\label{sec4_3}
\noindent
This subsection compares the four DoE enrichment methods:  MoV, t-LF, g-LF, and MD. We perform a similar statistical analysis as the previous subsection with respect to accuracy and efficiency metrics. Each DoE enrichment method is evaluated based on 48 cases (16 examples $\times$ 3 types of surrogates).

\begin{figure}[ht]
   	\begin{center}
	\includegraphics[width=10cm]{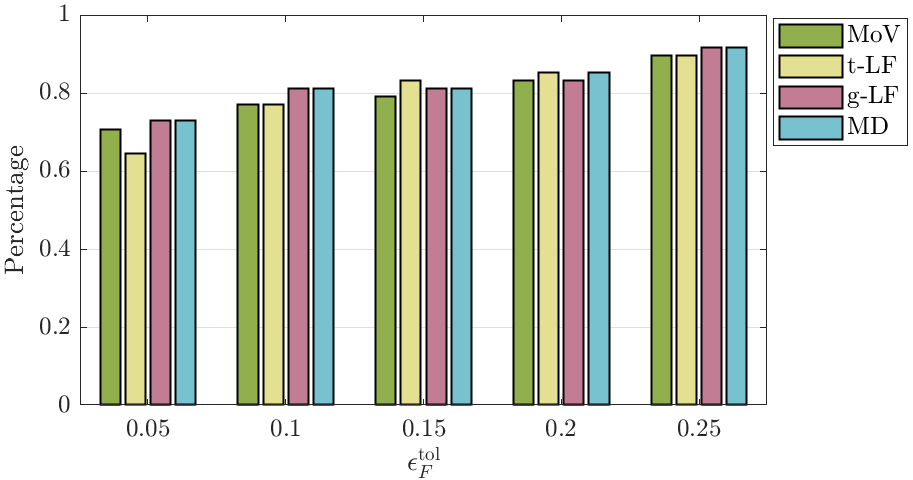}
	\caption{Comparison of DoE enrichment methods in terms of accuracy. 
\textit{ The accuracy metric uses the percentage of  cases (out of the 48 cases) that the DoE enrichment method can converge (i.e., $\epsilon_{F} \le \epsilon^{\text{tol}}_{F}$) within a fixed maximum number of iterations}. }
	\label{fig6_1}
	\end{center}
\end{figure}

As for the accuracy metric, Fig.\ref{fig6_1} shows the percentage of times that a method can converge within the pre-specified maximum number of iterations. This percentage increases with $\epsilon_{F}^{\text{tol}}$. However, for a fixed threshold $\epsilon^{\text{tol}}_{F}$, the variation among the 4 enrichment methods is smaller than that of the 3 surrogate models in Fig.\ref{fig5_1}. Therefore, we conclude that the accuracy metric is dominated by the surrogate module rather than the DoE enrichment module. In addition, the difference between  AL strategies (MoV, t-LF, and g-LF) and uniform design (MD) is surprisingly minor, which implies that in most cases, AL approaches may not significantly improve accuracy compared to uniform designs.

\begin{figure}[ht]
   	\begin{center}
	\includegraphics[width=10cm]{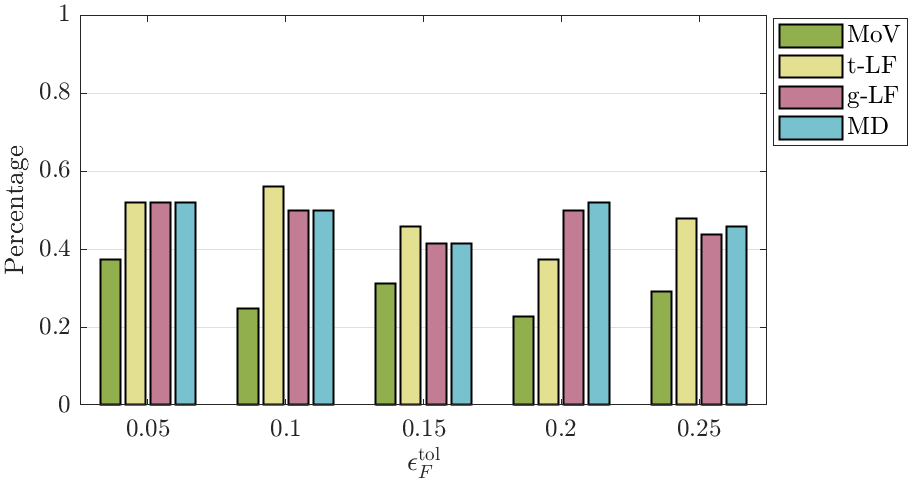}
	\caption{Comparison of DoE enrichment methods in terms of efficiency. \textit{ The efficiency metric uses the percentage of  cases (out of the 48 cases) that the DoE enrichment method requires the minimum iterations to converge  (i.e., $\epsilon_{F} \le \epsilon^{\text{tol}}_{F}$) }.}
	\label{fig6_2}
	\end{center}
\end{figure}

Fig.\ref{fig6_2} provides an assessment regarding the efficiency metric, exhibiting irregular fluctuations in contrast to Fig.\ref{fig5_2}. It is observed that t-LF is generally the most efficient except  when $\epsilon_{F}^{\text{tol}}=0.2$, while MoV is the least efficient method overall. Besides, the most surprising result is that MD is comparable to AL approaches, it even performs the best when $\epsilon_{F}^\text{tol}=0.2$. This result again reveals that AL approaches can not overwhelmingly outperform uniform designs regarding efficiency. Furthermore, both Fig.\ref{fig6_1} and Fig.\ref{fig6_2} show that the t-LF which integrates information both from the surrogate model and input PDF cannot consistently outperform g-LF which only utilizes the information from the surrogate model. In fact, when  $\epsilon_{F}^{\text{tol}}$ is set as 0.05, 0.1, or 0.25, t-LF gives a better/comparable performance with respect to efficiency in  Fig.\ref{fig6_2} but exhibits worse performance with respect to accuracy in  Fig.\ref{fig6_1}. When $\epsilon_{F}^{\text{tol}}=0.20$, g-LF outperforms t-LF. Therefore, t-LF does not consistently outperform g-LF.

\begin{figure}[ht]
   	\begin{center}
	\includegraphics[width=16cm]{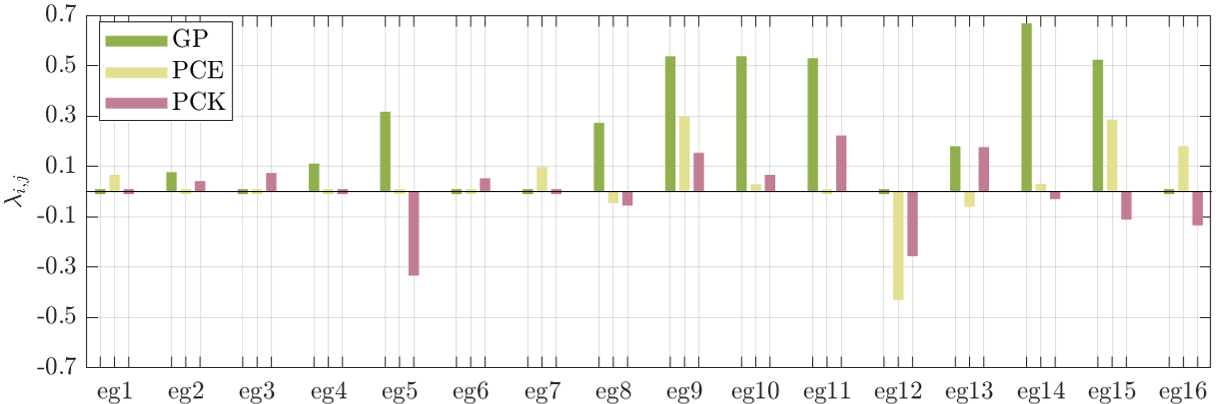}
	\caption{ $AL$ versus $uniform$ design on constructing global surrogates with  $\epsilon_{F}^{tol}=0.1$.  \textit{Positive $\lambda_{i,j}$  denotes AL is effective and negative value represents the opposite.}}
	\label{fig7_1}
	\end{center}
\end{figure}

For a detailed comparison,  we further analyze the results in Fig.\ref{fig6_2} by focusing on $\epsilon_{F}^{\text{tol}}=0.1$. The analysis measures the effect of \textit{AL versus uniform design} for constructing global surrogates. To do this, we split the data into two sets: one set refers to the average number of model evaluations $N_{\mathcal{M}_{i,j}}^{\rm{uniform}}$ from the uniform design, and another set refers to $N_{\mathcal{M}_{i,j}}^{\rm{AL}}$ from AL approaches, where the subscript denotes the results computed by $j^{th}$ type of surrogate model (GP, PCE and PCK) in the $i^{th}$ example. Since we have 3 AL strategies, $N_{\mathcal{M}_{i,j}}^{\rm{AL}}$ refers to the most efficient results among MoV, t-LF, and g-LF. Therefore, a novel metric is proposed to measure the benefit of using AL strategies, i.e.,
\begin{equation}
\label{eq_sec4_eq3}
 \lambda_{i,j} = \frac{ N_{\mathcal{M}_{i,j}}^{\rm{uniform}} - N_{\mathcal{M}_{i,j}}^{\rm{AL}} } { N_{\mathcal{M}_{i,j}}^{\rm{uniform}} }.
 \end{equation}
The positive $\lambda_{i,j}$ implies that the AL approach is more efficient than MD, and the negative represents the opposite. Fig.\ref{fig7_1} illustrates the metric $\lambda_{i,j}$ over 16 benchmarks solved by 3 different surrogates. Here the results again illustrate that the AL method is not always superior to the uniform design. The signs of the indicator $\lambda_{i,j}$ vary with the problems and the adopted surrogates. In all 16 examples, GP using AL methods performs the best within 10 cases, but this number drops to 7 for PCE and PCK. Besides, the improved efficiency for GP ranges from $0\% \sim 67\%$ in all examples. But this percentage can drop to be negative for PCE ($-43\% \sim 30\%$) and PCK ($-33\% \sim 22\%$). Hence, GP is more likely to obtain a greater improvement from AL approaches while PCE and PCK are less possible. Furthermore, the benefit from using AL is small compared with the reported results (e.g., see  \cite{Wang2020,crombecq2011novel}) that the improved effect ranges in the order of $1 \sim 100$ times. Meanwhile, the negative effect of adopting AL should not be ignored. To figure out this phenomenon, we will provide an additional investigation in Section \ref{sec5}.

\subsection{Stopping criteria settings}
\label{sec4_4}
\noindent
This subsection focuses on studying the parameter settings for the two stopping criteria. Specifically, we investigate the best settings of the stopping criteria thresholds and the number of consecutive triggers for stopping conditions. This is achieved by performing a statistical analysis over the $1,920$ UQ analyses in terms of efficiency and cost metrics.  The efficiency metric, similar to that in the previous two subsections, is evaluated on each run. The cost metric is defined as the mean value of the normalized number of model evaluations. The normalized results are obtained by dividing the number of model evaluations by  $N^{\min}_{\mathcal{M}}$ (i.e., the minimum number of model evaluations required to reach $\epsilon_{F} \le \epsilon^{\text{tol}}_{F}$). Unless otherwise stated, the error tolerance $\epsilon^{\text{tol}}_{F}$ is taken as 0.1 in this subsection.
 
\begin{figure}[ht]
   	\begin{center}
	\includegraphics[width=14cm]{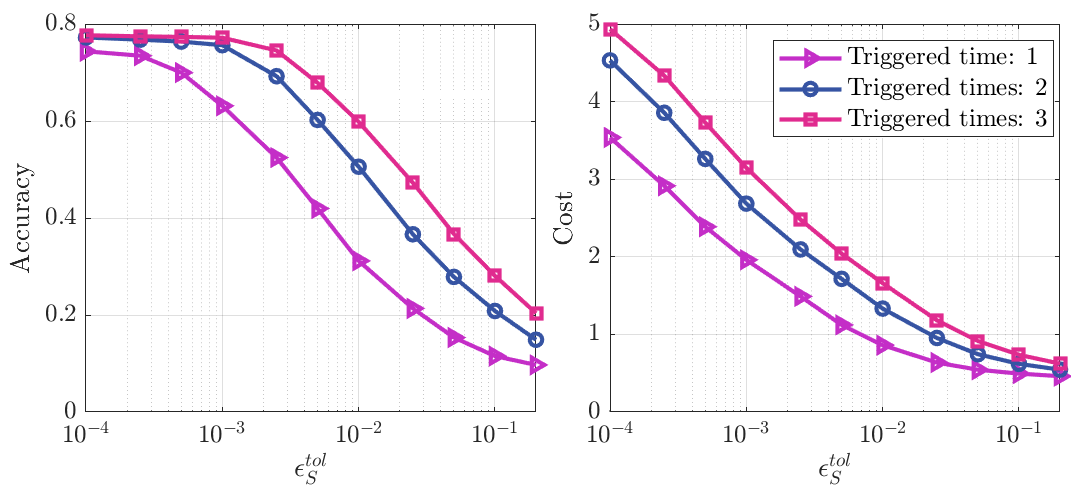}
	\caption{The influence of static-based stop criterion parameters on the accuracy and efficiency metric.}
	\label{fig9_1}
	\end{center}
\end{figure}

\begin{figure}[ht]
   	\begin{center}
	\includegraphics[width=14cm]{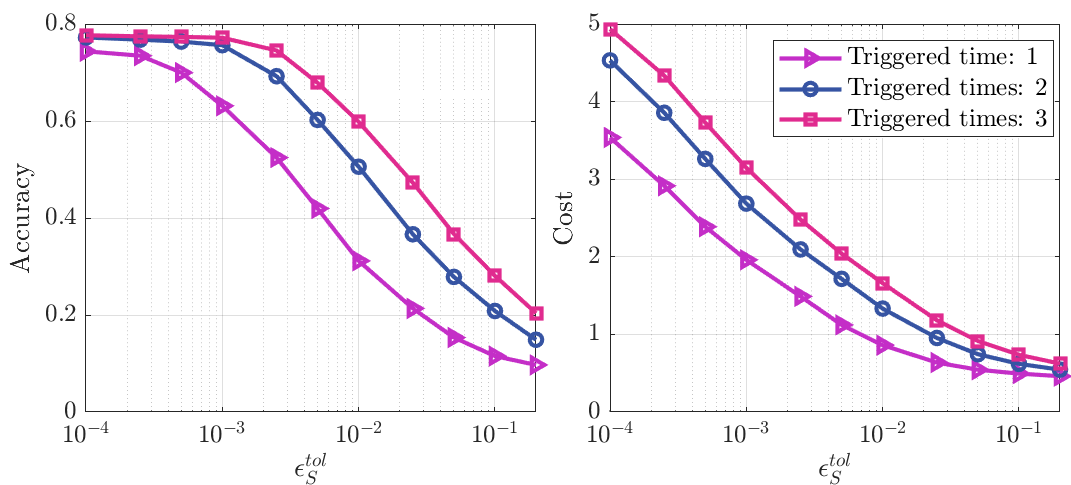}
	\caption{The influence of variance-based stop criterion parameters on the accuracy and efficiency metric.}
	\label{fig10_1}
	\end{center}
\end{figure}

The DoE enrichment process is terminated when the stopping criterion (${\epsilon}_{S}  \leq \epsilon^{\text{tol}}_{S}$ or ${\epsilon}_{V}  \leq \epsilon^{\text{tol}}_{V}$ ) is triggered within a given number of consecutive iterations. We consider a series of values for the thresholds (i.e., $\epsilon^{\text{tol}}_{S}$ and $\epsilon^{\text{tol}}_{V}$)  and set the consecutive triggered time(s) as 1, 2, and 3. With different thresholds and the number of consecutive triggers for stop conditions, we respectively compute the accuracy and cost metrics. The results are provided in Fig.\ref{fig9_1} for static-based criteria and in Fig.\ref{fig10_1} for variance-based criteria. Both stopping criteria show a monotonous pattern in which both accuracy and cost metrics decrease with the increased thresholds. In addition, for a fixed error tolerance, increasing the number of consecutive triggers for stopping conditions leads to more accurate results at the expense of higher cost. 

\begin{figure}[ht]
     \centering
     \begin{subfigure}[b]{0.4375\textwidth}
         \centering
         \includegraphics[width=\textwidth]{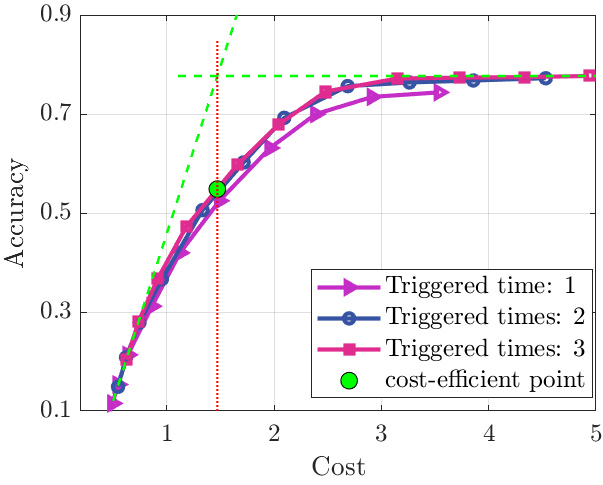}
         \caption{Static-based stopping criteria $\epsilon_{S}$}
         \label{fig9_10_sub1}
     \end{subfigure}
     \hfill
     \begin{subfigure}[b]{0.4375\textwidth}
         \centering
         \includegraphics[width=\textwidth]{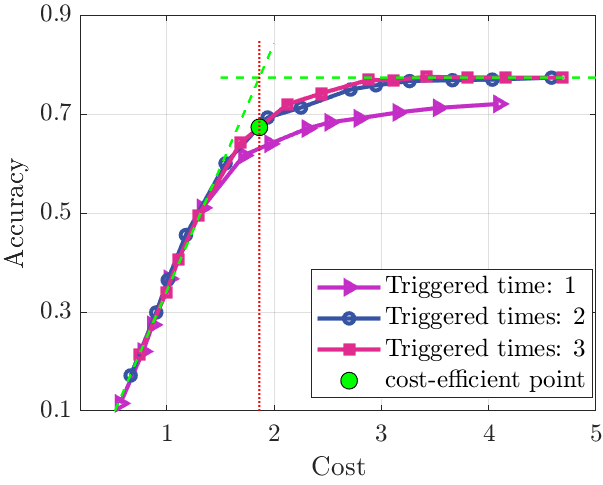}
         \caption{Variance-based stopping criteria $\epsilon_{V}$}
         \label{fig9_10_sub2}
     \end{subfigure}
        \caption{Cost-Accuracy curve on training sequential surrogates with different settings of two types of stopping criteria.}
        \label{fig9_10}
\end{figure}

To reveal more details, we combine the Accuracy-axis and Cost-axis of two criteria into one figure, as shown in Fig.\ref{fig9_10}. Firstly, we compare the  effects of threshold settings of two stopping criteria. The same trends can be found in Fig.\ref{fig9_10_sub1} and \ref{fig9_10_sub2}. For both stopping criteria, there exists a cost-efficient point beyond which increasing the cost does not significantly improve accuracy. However, some differences between the two stopping criteria can be observed. In the beginning, the accuracy grows almost linearly with respect to the increased cost. The initial slope for the static-based criterion is larger than that of the variance-based criterion. As the cost increases, the slope of the static-based criterion reduces faster than that of the variance-based criterion. Next, we compare the effect of the prescribed number of consecutive triggers for both stopping criteria. It is observed from Fig.\ref{fig9_10} that their effects are marginal when the accuracy metric is smaller than 0.6. Beyond this point, setting the number to 2 or 3 is more cost-efficient than setting it to 1. Therefore, we recommend setting the number of consecutive triggers to be larger than 1 if the error threshold is relatively tight.

To balance the cost and accuracy, we introduce the computational details on the \textit{cost-efficient point}, highlighted in Fig.\ref{fig9_10}, to determine the ``best'' thresholds for the stopping criteria. First, we define a \textit{cutoff cost} as the position where the initial and final slope of the Cost-Accuracy curve meet. The initial slope refers to the tangent line through the starting point of the curve, while the final slope corresponds to the horizontal line reaching maximum accuracy. Then, the projection of the cutoff cost on the Cost-Accuracy curve is defined as the cost-efficient point.  The coordinates of the cost-efficient points are $(1.47, 0.548)$ for $\epsilon_S$ in Fig.\ref{fig9_10_sub1} and $(1.86, 0.673)$ for $\epsilon_V$ in Fig.\ref{fig9_10_sub2}. These two coordinates suggest that using the criterion $\epsilon_V$ produces accurate estimations but requires more cost. In contrast, using the criterion $\epsilon_S$ needs less computational costs but obtains less precise results.

Based on the cost-efficient points, we can locate the balanced thresholds on the curves in Fig.\ref{fig9_1} and Fig.\ref{fig10_1}. This procedure can be generalized to other error tolerance values. In Table \ref{tab1}, we give the recommended thresholds for the two stopping criteria based on the concept of cost-efficient points.

\begin{table}[ht]
\renewcommand\arraystretch{1.5}
  \centering
  \caption{Recommended thresholds for different configurations of the stopping criteria}
  \label{tab1}
\begin{tabular}{ccccc}
\toprule
\multirow{2}{*}{$\epsilon^{\text{tol}}_{F}$} & \multicolumn{2}{c}{consecutive triggers: 2}                   & \multicolumn{2}{c}{consecutive triggers: 3}                   \\ \cline{2-5} 
                                             & $\epsilon^{\text{tol}}_{S}$ & $\epsilon^{\text{tol}}_{V}$ & $\epsilon^{\text{tol}}_{S}$ & $\epsilon^{\text{tol}}_{V}$ \\ [0.5ex]
                                             \midrule
0.05                                         & 0.002                       & 0.071                       & 0.004                       & 0.085                       \\
0.1                                          & 0.008                       & 0.225                       & 0.016                       & 0.274                       \\
0.15                                         & 0.014                       & 0.320                       & 0.023                       & 0.390                       \\
0.2                                          & 0.017                       & 0.441                       & 0.028                       & 0.511                       \\
0.25                                         & 0.020                       & 0.460                       & 0.034                       & 0.533                       \\ \bottomrule
\end{tabular}
\end{table}

\begin{figure}[ht]
   	\begin{center}
	\includegraphics[width=16cm]{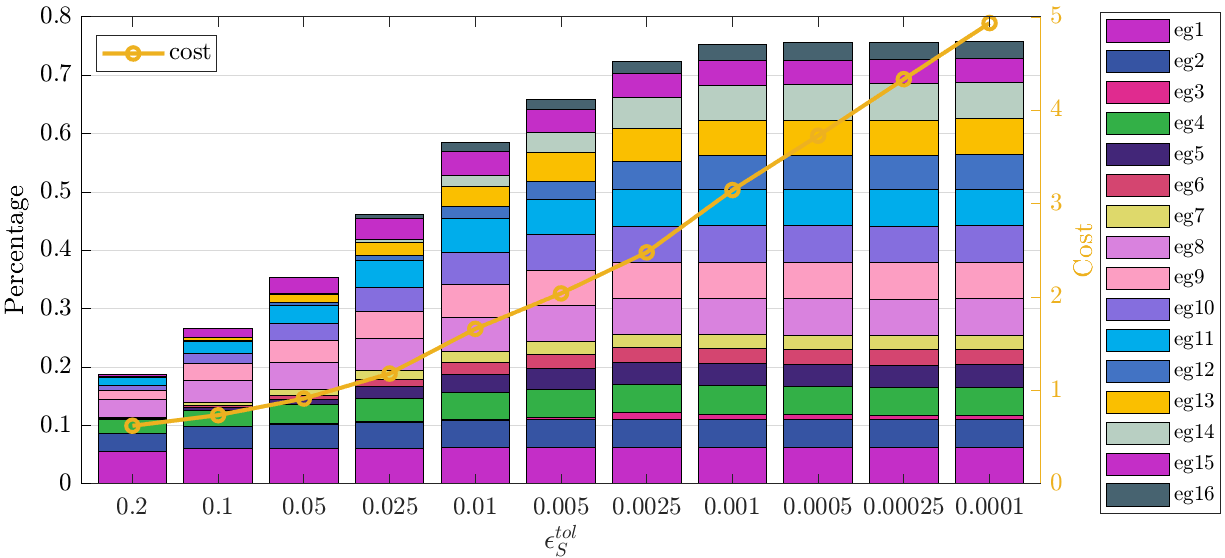}
	\caption{Decomposition of the aggregated accuracy metric into individual contributions from the 16  examples.}
	\label{fig12}
	\end{center}
\end{figure}

To better understand the convergence trend across different examples, the accuracy curve (triggered time: 3) in Fig.\ref{fig9_1} is decomposed by detailed contributions from each example. The total accuracy represented in percentage is equal to the sum of the convergence rates (i.e., the proportion of cases that  $\epsilon_{F} \le \epsilon^{\text{tol}}_{F}$) of the 16 examples. The aggregated accuracy metric together with the total cost metric is shown in Fig.\ref{fig12}. Here, we observe that the optimal stopping threshold varies from example to example. For instance, Example \#1 can converge accurately by setting a large threshold as 0.2. Examples \#2, \#4, and \#8 $\sim$ \#11 can achieve high accuracy when the threshold is adjusted to 0.01. Most problems reach their maximum precision when tuning the threshold as 0.0025. After this value, the accuracy rises slowly with increasing the cost. For Examples \#3, \#6, \#7, and \#16, reaching high accuracy is hindered by the slow convergence rate. In this case, two solutions can improve the accuracy: (\textit{i}) increase the maximum number of model evaluations, and (\textit{ii}) change the surrogate settings (e.g., other types of surrogates, different kernel functions, higher PCE degrees, etc.). In summary, the optimal setting for the stopping criteria depends on the problem.

\section{Discussion and Recommendation}
\label{sec5}
\subsection{AL versus uniform design}
\label{sec5_1}
\noindent
In section \ref{sec4_3}, we compared the construction of global surrogates to computing full distribution when using three active learning  (AL) designs versus a uniform design.  Three findings emerge:
\begin{itemize}
\item  the profitability of  using AL is lower than expected;
\item  uniform design can outperform AL;
\item  the surrogate model GP benefits more from AL than PCE and PCK.
\end{itemize}
This subsection discusses each finding and offers detailed explanations.

For the first finding, we attribute it to the fact that the maximum profitability of using AL is bounded by the discrepancy of local nonlinearity (DoLN)  in the simulators.\footnote{Here, the local nonlinearity can be measured by the highest order of Taylor series expansions required to accurately substitute the simulator around a specific local point. Higher orders imply  greater local nonlinearity. } A low DoLN implies a nearly uniform distribution of local nonlinearity across the design space, regardless of the magnitude of the average nonlinearity; while a high DoLN implies significant variations of local nonlinearity. Therefore, AL techniques are effective in cases of high DoLN, where they allocate the samples over the design space proportional to the local nonlinearity. However, they become inefficient in cases of low DoLN. For example, in our benchmark study, Example \#4 involves a simple simulator consisting of two symmetric 2-dimensional linear functions, where an arbitrary DoE (with more than three non-colinear points) suffices for each linear function. Conversely, Example \#6, characterized by a strongly nonlinear function with widespread local maxima, requires a densely and uniformly distributed set of samples for an accurate surrogate.  These two examples have different levels of nonlinearity, but their DoLNs are both low. Our observation is that all the 16 benchmark examples, despite being collected from various  literature, exhibit a low level of DoLN, resulting in the lower-than-expected profitability from using AL. 

The second finding can be attributed to the inaccurately estimated local errors from surrogates, which in turn influences the selection of training points and leads to reduced learning efficiency.  In fact, local error estimations rely on certain assumptions. For example, the variance-based error of GP assumes that the underlying simulator follows a Gaussian process. Cross-validation error assumes that the surrogate predictions are sensitive to the removal  of a proportion of samples from the DoE. As highlighted in \cite{jin2002sequential},  if the estimated local errors are misleading, the iterative surrogates might diverge from the real simulators. Although learning functions can, to some extent, correct this divergence, redundant points are inevitably added in comparison to using accurate local errors. Therefore, uniform design can outperform AL,  especially when the DoLN is low.

The third finding is attributed to the use of different basis functions for the three surrogates:  GP, PCE, and PCK. Equation \eqref{eq_sec2_2} indicates that the  three surrogates comprise either a global predictor $\hat{\mathcal{M}}^\mathrm{G}({\boldsymbol{x}})$ or a local predictor $\hat{\mathcal{M}}^\mathrm{L}({\boldsymbol{x}})$. Both two predictors share a similar form, i.e., the dot product of  basis functions and deterministic coefficients, but their basis functions have distinct features.  The basis function $R(\boldsymbol{x},\boldsymbol{x}_j^d|\boldsymbol{\theta})$ of the local predictor relies on the position of the training points $\boldsymbol{x}_j^d$. Besides, its influence is localized at the neighbor of  $\boldsymbol{x}_j^d$, with the correlations decay as $\boldsymbol{x}$ moves away from the training points. This feature makes $\hat{\mathcal{M}}^\mathrm{L}({\boldsymbol{x}})$ adept at capturing local variations.  On the other hand, the polynomial basis functions ${\boldsymbol{\Psi}(\boldsymbol{x})}$ of the global predictor span over the design space,  making $\hat{\mathcal{M}}^\mathrm{G}({\boldsymbol{x}})$ better suited for capturing global trends. These two distinct features determine that   $\hat{\mathcal{M}}^\mathrm{G}({\boldsymbol{x}})$ and $\hat{\mathcal{M}}^\mathrm{L}({\boldsymbol{x}})$ require different distribution patterns of  design samples to achieve the same level of approximation accuracy. Specifically, $\hat{\mathcal{M}}^\mathrm{L}({\boldsymbol{x}})$ requires more samples in the nonlinear region but fewer in the linear region. However,  $\hat{\mathcal{M}}^\mathrm{G}({\boldsymbol{x}})$ is globally fitted to capture the underlying patterns in data. For a pure global predictor, a sufficient number of samples (comparable to the number of free regression parameters $\boldsymbol{{\beta}}$) is more crucial than searching for the optimal allocations of the training samples.

 AL techniques specialize in searching regions with large local prediction error/uncertainty,  and these regions are generally nonlinear. Therefore, the  AL  is more effective in accelerating the training of  local predictors. As a result, the benefits of using AL  are more pronounced for the pure local-prediction-based GP  than for the global-prediction-based PCE.

\begin{figure}[ht]
   	\begin{center}
	\includegraphics[width=12cm]{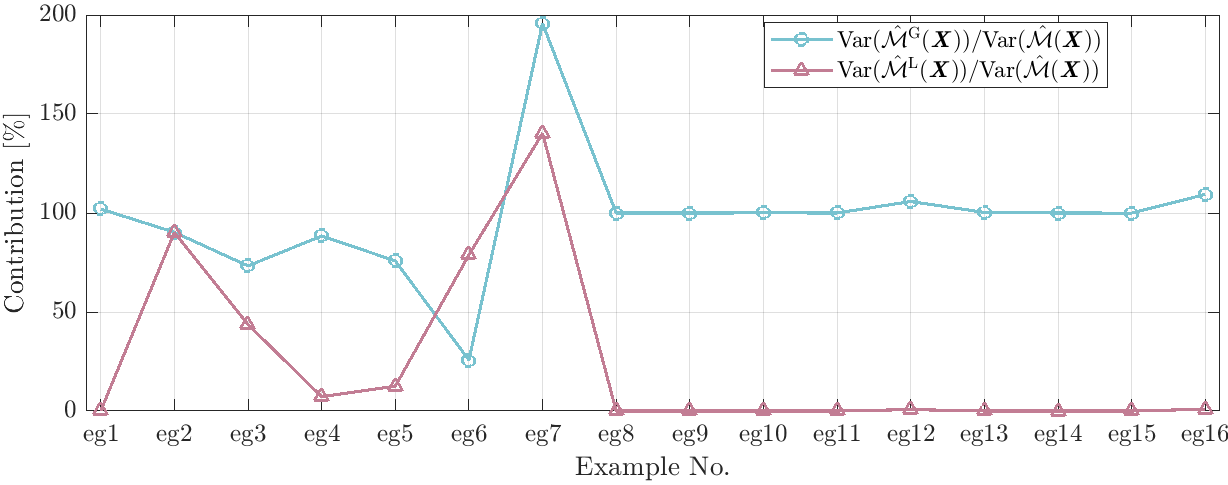}
	\caption{Contributions of the global and local predictors on the total variance of surrogate model predictions.}
	\label{fig13}
	\end{center}
\end{figure}

To understand why PCK benefits less from AL than GP, we conduct variance decomposition for PCK to quantify the explained variability separately from its global and local predictors. Specifically, we apply the variance operator to both sides of Eq.\eqref{eq_sec2_2} with respect to random input $\boldsymbol{X}$, obtaining: 
\begin{equation}
\label{eq_sec5_2}
\begin{aligned}
\text{Var}\left(  \mathcal{\hat{M}}({\boldsymbol{X}})  \right)   = &\text{Var}\left(\hat{\mathcal{M}}^\text{G}(\boldsymbol{X})\right) + \text{Var}\left(\hat{\mathcal{M}}^\text{L}(\boldsymbol{X})\right)\\ & + 2 \text{CoV}\left(\hat{\mathcal{M}}^\text{G}(\boldsymbol{X}),\hat{\mathcal{M}}^\text{L}(\boldsymbol{X})\right),
\end{aligned}
\end{equation}
where the left-hand side denotes the total variance, and the right-hand side separately denotes the explained variance by $\hat{\mathcal{M}}^\text{G}(\boldsymbol{x})$ and $\hat{\mathcal{M}}^\text{L}(\boldsymbol{x})$ in addition to  their covariance.   We compute the sample variance and covariance in Eq.\eqref{eq_sec5_2} using $10^5$ random samples.  We illustrate this decomposition for the 16 benchmark examples and train the PCK using the two-step learning function until reaching the maximum DoE size (as prescribed in Section \ref{sec4_1_2}). Fig.\ref{fig13} illustrates the variance contributed from the global predictor and local predictor, depicting their percentages relative to the total variance. It is observed that only in Examples \#2 $\sim$ \#7 the local predictors make  noticeable contributions to the total variation. For all the other examples, the global predictor dominates.  Therefore, the PCK degenerates to PCE for most of the examples studied. This finding can also explain why GP is benefited more from AL than PCK. 


\subsection{Recommendations to construct surrogates}
\label{sec5_2}
\noindent
In this section, we summarize the findings to provide recommendations on constructing surrogates for computing probability distributions. Our suggestions are set up based on the statistical analysis of the results for solving the 16 benchmark problems. The conclusions might become inappropriate when practitioners study problems that are significantly different from the benchmarks. Our general finding is that there is no one strategy that is consistently superior to the others. The best strategy is correlated to features of the problem (e.g., discrepancy of local nonlinearity), the chosen types of surrogates, the expected accuracy requirement, and the computational budget. The recommendations are listed in the following. 

\begin{enumerate}
\item \textbf{Surrogate module}: In general, we recommend practitioners to use PCK, which combines the advantages  of global approximation from PCE and the property of exact interpolation from GP. Additionally, we suggest setting the trend of PCK with a low-degree PCE.  This recommendation is based on our finding that the simulators in our benchmarks, despite being selected from a diverse literature,  exhibit a low-degree polynomial structure. However, it is also important to note that PCK cannot consistently outperform GP and PCE. The best choice between these models ultimately depends on the problem  and the required level of accuracy. 
\item \textbf{DoE enrichment module}: For PCE and PCK, we suggest using uniform design, as the improvement of using AL is generally limited, and AL can sometime perform worse than uniform design. For GP,  AL techniques are recommended, although we cannot identify the best AL technique.  Our study shows that choosing the proper type of surrogate is more crucial and can result in saving a greater number of training points.  When AL techniques are preferable in practice, we recommend consulting the latest review papers about AL techniques for GP \cite{liu2018survey,fuhg2021state}. However, it is worth noting that the best choice of AL methods for GP often depends on the specific features of the problem \cite{fuhg2021state}. 
\item   \textbf{Stopping criteria module}: First, we recommend setting the thresholds provided in Table \ref{tab1} for the two stopping criteria, i.e.,  static-based criterion $\epsilon_S$ and variance-based criterion $\epsilon_V$.  These thresholds are determined by balancing the accuracy and cost.  Additionally, we suggest choosing the stopping criterion $\epsilon_V$ when the computational budget is more relaxed or higher level of accuracy is required. Conversely, we suggest choosing the static-based criterion $\epsilon_S$ in scenarios where computational resources are limited or a lower level of accuracy is acceptable. Finally, we recommend setting the number of consecutive triggers of stopping criteria to a number larger than 1.

\end{enumerate}

\section{Conclusion}
\label{sec6}
\noindent
The paper conducts a comprehensive comparative study on constructing surrogate models for full distribution estimations. It addresses two primary objectives: (\textit{i}) compare uniform design to active learning design, and (\textit{ii}) provide recommendations for practitioners on constructing surrogate models. The study proposes a modular framework that comprises three essential modules: types of surrogate, DoE enrichment methods, and stopping criteria. Various representative methods are investigated for each module to provide insights into their comparative performance. 

Our first conclusion is that active learning techniques do not consistently provide a systematic improvement compared to uniform design. We  provide three reasons (in Section \ref{sec5_1})  to explain why the benefit of using AL is lower than expected and varies for different surrogate models. Our second conclusion is that no strategy can consistently outperform the others in constructing   surrogates, and the recommended configurations (in Section \ref{sec5_2})  depend on the chosen types of surrogate model, features of the problem (especially the discrepancy of the local nonlinearity of the simulators), and the trade-off between target accuracy and affordable cost.

Although the focus of the study is on constructing surrogate models for  probability distribution estimation, the conclusions are expected to be relevant to other applications such as sensitivity analysis, parameter estimation, and reliability-based optimization.

Finally, this paper proposed a novel pool-based maximin-distance design method that exhibits promising performance in estimating distribution functions. Although this new uniform design method is used only as a baseline for the comparative study, it is expected that the method could be promising for other applications.

\section*{Acknowledgments}
The first and third authors are funded by the Italian Ministry of Education, University and Research (MIUR) in the frame of the ``Departments of Excellence 2023-2027'' (grant L 232/2016). The last author is by funded the European Union under Next GenerationEU. PRIN 2022 Prot. n 2022MJ82MC\_001.

\appendix
\section{Supplementary materials}
\label{Appedndix_A}
Supplementary materials of this manuscript are additionally provided.
\bibliographystyle{elsarticle-num}
\bibliography{MyRefs}

\end{document}


\begin{frontmatter}



\title{Supplementary materials of ``Active learning based surrogate-aided methods for full probability distribution estimation: uniform or adaptive design?''}


\author[inst1]{Maijia Su}
\affiliation[inst1]{organization={Department of Civil, environmental and mechanical engineering, University of Trento, Trento},
            country={Italy}}           
\author[inst2]{Ziqi Wang}
\author[inst1]{Oreste Salvatore Bursi}
\author[inst1]{Marco Broccardo}
\affiliation[inst2]{organization={Department of Civil and Environmental Engineering, University of California, Berkeley},
            country={United States}}



\end{frontmatter}


\section{Benchmark Problems}
This section introduces 16 benchmark examples, including 13 analytical examples (\#1 $\sim$ \#13) and 3 structural examples based on FEMs (\#14 $\sim$ \#16). The performance functions are provided in Table \ref{table:1}, and the probabilistic models together with the ranges of quantities of interest (QoI), i.e., $[y_{\min},y_{\max}]=[F^{-1}_Y(10^{-3}), F^{-1}_Y(1-10^{-3})]$, are presented in Table \ref{table:2}. The bounds  $y_{\min}$ and $y_{\max}$ are estimated by crude MCS with $10^6$ samples. Benchmarks \#14 and \#16 are two widely studied examples in the context of structural reliability. As for these two examples, this study adopts the same structural simulators and uncertainty models as the original problems, but we focus on computing the full probability distributions of the structural responses of interest rather than the failure probabilities. 

Benchmark\#15 studies a 10-story steel structure developed in \cite{ohtori2004benchmark} for vibration control problems. We adopt the simplified story-shear model of this structure. The model is assumed to be linearly elastic and  has 5\% Rayleigh damping for the first and second modes. In addition, the structure is subjected to a real ground motion, i.e., the N-S component of the 1940 El-Centro earthquake. The story parameters of this story-shear model are presented in Table \ref{table:3}. The probabilistic model supposes that the story masses and story stiffness follow the independent logarithmic normal distributions. The mean values of the random variables are taken as the corresponding real parameters as shown in Table \ref{table:3}, and the coefficients of variation are set as 0.02 for each story mass and 0.05 for each story stiffness. The total number of random variables of this benchmark is 20. We perform the analysis in MATLAB and specify the maximum story drift as the response of interest.

\begin{longtblr}[
    caption = {Performance functions of the 16 Benchmarks}, 
    label = {table:1}, 
    note{1} = {Dim. refers to the dimensions of the problems}, 
  ]{
    colspec = {X[1,c]X[12,c]X[1,c]X[2,c]}, 
    rowhead = 1, 
  }
\hline
No. & Function descriptions & Dim. & Reference \\ [0.5ex]
\hline\hline
 1 &  $\mathcal{M}(\mathbf{X})=2.5-0.2357 \left( {{X}_{1}}-{{X}_{2}} \right)+0.00463 {{\left( {{X}_{1}}+{{X}_{2}}-20 \right)}^{4}}$ & 2 & \cite{grooteman2011adaptive} \\ 
\hline
 2 & $\mathcal{M}(\mathbf{X})=\sin \left( \frac{5 {{X}_{1}}}{2} \right)+2-\frac{\left( X_{1}^{2}+4 \right) \left( {{X}_{2}}-1 \right)}{20}$ & 2 & \cite{schueller2004critical} \\ 
\hline 
3 & $\mathcal{M}(\mathbf{X})=\min \left\{ \begin{array}{*{35}{l}}
   3+0.1{{\left( {{X}_{1}}-{{X}_{2}} \right)}^{2}}-\frac{\left( {{X}_{1}}+{{X}_{2}} \right)}{\sqrt{2}};  \\
   3+0.1{{\left( {{X}_{1}}-{{X}_{2}} \right)}^{2}}+\frac{\left( {{X}_{1}}+{{X}_{2}} \right)}{\sqrt{2}};  \\
   \left( {{X}_{1}}-{{X}_{2}} \right)+\frac{k}{\sqrt{2}};\left( {{X}_{2}}-{{X}_{1}} \right)+\frac{k}{\sqrt{2}}  \\
\end{array} \right\}$
 & 2 & \cite{Echard2011} \\ 
\hline
 4 & $\mathcal{M}(\mathbf{X})=\min \{{{X}_{1}}-{{X}_{2}};{{X}_{1}}+{{X}_{2}}\}$ & 2 & \cite{Wang2020} \\ 
\hline
 5 & $\mathcal{M}(\mathbf{X})=\min \left\{ \begin{matrix}
   2-{{X}_{2}}+\exp \left( -X_{1}^{2}/10 \right)+{{\left( {{X}_{1}}/5 \right)}^{4}};  \\
   {{3}^{2}}/2-{{X}_{1}}{{X}_{2}}  \\
\end{matrix} \right\}$
 & 2 & \cite{au1999new} \\ 
 \hline
 6 & $\mathcal{M}(\mathbf{X})=10-\sum\limits_{i=1}^{2}{\left( X_{i}^{2}-5\cos \left( 2\pi {{X}_{i}} \right) \right)}$ & 2 & \cite{Echard2011} \\ 
\hline
 7 & $\mathcal{M}(\mathbf{X})=\sin \left( {{X}_{1}} \right)+a{{\sin }^{2}}\left( {{X}_{2}} \right)+bX_{3}^{4}\sin \left( {{X}_{1}} \right)$ & 3 & \cite{ishigami1990importance} \\ 
\hline
 8 & $\mathcal{M}(\mathbf{X})={{X}_{1}}-\frac{32}{\pi X_{2}^{3}}\cdot \sqrt{\frac{X_{3}^{2}X_{4}^{2}}{16}+X_{5}^{2}}$ & 5 & \cite{schueller2004critical} \\ 
\hline
 9 & $\mathcal{M}\left( {{c}_{1}},{{c}_{2}},m,r,{{t}_{1}},{{F}_{1}} \right)=3r-\left| \frac{2{{F}_{1}}}{m\omega _{0}^{2}}\sin \left( \frac{\omega _{0}{{t}_{1}}}{2} \right) \right|$ & 6 & \cite{Echard2011} \\
\hline
 10 & $\mathcal{M}(\mathbf{X})=1-\frac{\sqrt{3\left( 1-{{0.3}^{2}} \right)}}{\pi {{X}_{1}}X_{2}^{2}{{\cos }^{2}}{{X}_{3}}} \left( \frac{{{X}_{6}}}{0.66}+\frac{{{X}_{5}}}{0.41{{X}_{4}}} \right)$ & 6 & \cite{shabakhty2011structural} \\ 
\hline
 11 & $\mathcal{M}(\mathbf{X})={{\left[ \frac{{{X}_{1}}{{X}_{2}}}{\frac{{{X}_{3}}{{\left( {{X}_{4}}\frac{2\pi }{60} \right)}^{2}}\left( X_{5}^{3}-X_{6}^{3} \right)}{3\times385.82\left( {{X}_{5}}-{{X}_{6}} \right)}} \right]}^{1/2}}-0.37473$ & 6 & \cite{wei2007structural} \\ 
\hline
 12 & $\begin{aligned}
   \mathcal{M}(\mathbf{X}) = & 15.59 {{10}^{4}}-\frac{{{X}_{1}} X_{2}^{3}}{2 X_{2}^{3}}\times \\ 
 & \frac{X_{4}^{2}-4 {{X}_{5}} {{X}_{6}} X_{7}^{2}+{{X}_{4}} \left( {{X}_{6}}+4 {{X}_{5}}+2 {{X}_{6}} {{X}_{7}} \right)}{{{X}_{4}} {{X}_{5}} \left( {{X}_{4}}+{{X}_{6}}+2 {{X}_{6}} {{X}_{7}} \right)} \\ 
\end{aligned}$ & 7 & \cite{schueller2004critical} \\ 
\hline
 13 & $\mathcal{M}(r_w,r,T_u,H_u,T_l,H_l,L,K_w)=\frac{2\pi {{T}_{u}}\left( {{H}_{u}}-{{H}_{l}} \right)}{\ln \left( r/{{r}_{w}} \right)\left( 1+\frac{2L{{T}_{u}}}{\ln \left( r/{{r}_{w}} \right)r_{w}^{2}{{K}_{w}}}+\frac{{{T}_{u}}}{{{T}_{l}}} \right)}$ & 8 & \cite{schobi2017rare}\\
\hline
 14 & A 2-dimensional 23 bars truss & 10 & \cite{blatman2010efficient}\\
\hline
 15 & A 10-floor shear structure & 20 & -\\
\hline
 16 & A 3-bays-5-floor Frame  & 21 & \cite{liu1991kiureghian} \\
\hline
\end{longtblr}

\begin{longtblr}[
    caption = {Probabilistic models and ranges of QoI of the 16 Benchmarks}, 
    label = {table:2}, 
    note{1} = {$\mathcal{N}$ denotes the normal distribution, $\mathcal{LN}$ denotes the log-normal distribution, $\mathcal{U}$ denotes the uniform distribution, $\mathcal{G}$ denotes the Gumbel distribution and $\mathcal{W}$ denotes the Weibull distribution.}, 
  ]{
    colspec = {X[1,m]X[10,c]X[3,c]}, 
    rowhead = 1, 
  }
\hline
No. & Probabilistic models  & $[y_{\min},y_{\max}]$ \\ [0.5ex]
\hline\hline
1 & $ \begin{matrix} 
{{X}_{1}}\sim \mathcal{N}(10,3),{{X}_{2}}\sim \mathcal{N}(10,3),{{X}_{2}}\sim \mathcal{N}(10,3) 
\end{matrix}$& $[12,2061]$ \\
\hline
2 & $ \begin{matrix} {{X}_{1}}\sim \mathcal{N}(1.5,1),{{X}_{2}}\sim \mathcal{N}(1.5,1) 
\end{matrix}$ & $[-1.71,3.41]$ \\
\hline
3 & $ \begin{matrix} {{X}_{1}}\sim \mathcal{N}(0,1),{{X}_{2}}\sim \mathcal{N}(0,1) 
\end{matrix}$ & $[-0.23,3.24]$ \\
\hline
4 & $ \begin{matrix} {{X}_{1}}\sim \mathcal{N}(0,1),{{X}_{2}}\sim \mathcal{N}(0,1) 
\end{matrix}$ & $[-4.67,2.63]$ \\
\hline
5 & $ \begin{matrix} {{X}_{1}}\sim \mathcal{N}(0,1),{{X}_{2}}\sim \mathcal{N}(0,1) 
\end{matrix}$ & $[-0.70,5.6]$ \\
\hline
6 & $ \begin{matrix} {{X}_{1}}\sim \mathcal{N}(0,1),{{X}_{2}}\sim \mathcal{N}(0,1)
\end{matrix}$ & $[-12.6,19.9]$ \\
\hline
7 & $ \begin{matrix} {{X}_{1}}\sim \mathcal{U}(-\pi,\pi),{{X}_{2}}\sim \mathcal{U}(-\pi,\pi), {{X}_{3}}\sim \mathcal{U}(-\pi,\pi)
\end{matrix}$ & $[-9.1,16.1]$  \\
\hline
8 & $\begin{matrix} 
     & {{X}_{1}}\sim \mathcal{U}(70,80), {{X}_{2}}\sim \mathcal{N}(39,0.1),{{X}_{3}}\sim \mathcal{G}(1342,272.9), \\
     & {{X}_{4}}\sim \mathcal{N}(400,0.1), {{X}_{5}}\sim \mathcal{N}(2.5\times {{10}^{5}},3.5\times {{10}^{4}}) \\
\end{matrix}$ & $[0.9,44.3]$ \\
\hline
9 & $\begin{matrix}
     & m\sim \mathcal{N}(1,0.05),{{c}_{1}}\sim \mathcal{N}(1,0.1),{{c}_{2}}\sim \mathcal{N}(0.1,0.01), \\ 
     & r\sim \mathcal{N}(0.5,0.05),{{F}_{1}}\sim \mathcal{N}(1,0.2),{{t}_{1}}\sim \mathcal{N}(1,0.2) \\ 
\end{matrix}$ & $[-0.43,1.41]$ \\
\hline
10 & $\begin{matrix}
      & {{X}_{1}}\sim \mathcal{N}\left( 7.0\times {{10}^{10}},3.50\times {{10}^{9}} \right), \\
      & {{X}_{2}}\sim \mathcal{N}\left( 2.50\times {{10}^{-3}},1.25\times {{10}^{-4}} \right), \\
      & {{X}_{3}}\sim \mathcal{N}(0.524,0.010480), 
        {{X}_{4}}\sim \mathcal{N}(0.90,0.0225), \\
      & {{X}_{5}}\sim \mathcal{N}\left( 8.0\times {{10}^{4}},6.4\times {{10}^{3}} \right), \\
      & {{X}_{6}}\sim \mathcal{N}\left( 7.0\times {{10}^{4}},5.6.0\times {{10}^{3}} \right) \\ 
\end{matrix}$ & $[0.39,0.725]$ \\
\hline
11 &  $\begin{matrix}
     & {{X}_{1}}\sim \mathcal{W}\left( 0.9377,4.59\times {{10}^{-2}} \right), \\ 
     & {{X}_{2}}\sim \mathcal{N}\left( 2.2\times {{10}^{5}},5.0\times {{10}^{3}} \right), \\ 
     & {{X}_{3}}\sim \mathcal{U}\left( 0.29,5.8\times {{10}^{-3}} \right), \\ 
     & {{X}_{4}}\sim \mathcal{N}\left( 2.1\times {{10}^{4}},1.0\times {{10}^{3}} \right), \\ 
     & {{X}_{5}}\sim \mathcal{N}(24.0,0.5), 
       {{X}_{6}}\sim \mathcal{N}(8,0.3) \\ 
\end{matrix}$ & $[-0.004,0.170]$ \\
\hline

12 & $\begin{matrix}
     & {{X}_{1}}\sim \mathcal{N}\left( 350,35 \right),{{X}_{2}}\sim \mathcal{N}\left( 50.8,5.08 \right), \\
     & {{X}_{3}}\sim \mathcal{N}\left( 3.81,0.381 \right), {{X}_{4}}\sim \mathcal{N}\left( 173,17.3 \right),\\ 
     & {{X}_{5}}\sim \mathcal{N}(9.38,0.938),{{X}_{6}}\sim \mathcal{N}(33.1,3.31), \\ 
     & {{X}_{7}}\sim \mathcal{N}(0.036,0.0036) \\ 
      \end{matrix}$ & $[-61008,143310]$ \\
\hline

13 & $\begin{matrix}
     & {{r}_{w}}\sim \mathcal{U}\left( 0.05,0.15 \right),r\sim \mathcal{LN}\left( 7.71,1.0056 \right),\\
     & {{T}_{u}}\sim \mathcal{U}\left( 63070,115600 \right),{{H}_{u}}\sim \mathcal{U}\left( 990,1100 \right),\\
     & {{T}_{l}}\sim \mathcal{U}(63.1,116),{{H}_{l}}\sim \mathcal{U}\left( 700,820 \right),\\ 
     & L\sim \mathcal{U}(1120,1680),{{K}_{w}}\sim \mathcal{U}(9855,12045) \\ 
\end{matrix}$ &  $[12,230]$ \\
\hline
14 & Refer to \cite{blatman2010efficient}  & $[5.2, 12.2]$ \\
\hline
15 & Defined in this material  & $[0.0121, 0.0174]$ \\
\hline
16 & Refer to \cite{liu1991kiureghian}  & $[0.007,0.041]$ \\
\hline
\end{longtblr}

\begin{table}[ht]
\caption{Parameters of the 10-story steel structure in Benchmark\#15}
\label{table:3}
\centering
\begin{tabular}{cccc}
\hline
Story & Height (m) & Story mass ($10^5 \mathrm{kg}$) & Story stiffness ($10^8 \mathrm{N/m}$) \\ [0.5ex]
\hline\hline
1     & 5.49       & 2.82                  & 1.007                       \\
2     & 3.96       & 2.76                  & 1.358                       \\
3     & 3.96       & 2.76                  & 1.338                       \\
4     & 3.96       & 2.76                  & 1.330                       \\
5     & 3.96       & 2.76                  & 1.228                       \\
6     & 3.96       & 2.76                  & 1.248                       \\
7     & 3.96       & 2.76                  & 1.237                       \\
8     & 3.96       & 2.76                  & 1.223                       \\
9     & 3.96       & 2.76                  & 1.208                       \\
10    & 3.96       & 2.76                  & 1.187                       \\
\hline
\end{tabular}
\end{table}

\section{Learning Functions}
This section introduces three learning functions, serving as a supplement to  Section 2.5 in the main text.

\subsection{MoV learning function}
This learning function selects the points based on the maximum of variance (MoV) criteria. The MoV learning function is defined as
\begin{equation}
\label{eq_1}
L_\text{MoV}(\boldsymbol{x}) = \sigma^2_{\hat{\mathcal{M}}}(\boldsymbol{x}),
\end{equation}
where $ \sigma^2_{\hat{\mathcal{M}}}(\boldsymbol{x})$ is the prediction error/variance measured by the surrogate GP, PCE, or PCK.

\subsection{Two-step learning function}
The two-step learning function was proposed in \cite{Wang2020}. Specifically, the first step focuses on identifying a position $y^*$ in the range $[y_{\min}, y_{\max}]$.  The position  $y^*$  is identified by maximizing a localized error $W_L(y^{\prime})$ centered at $y^{\prime}$.  This localized error is built by weighting the error metric in Eq.(8) in the main text (i.e., the errors formulated by the bounded surrogates $\hat{\mathcal{M}}^k(x)=\hat{\mathcal{M}}(x)+k\sigma_{\hat{\mathcal{M}}}(\boldsymbol{x})$) by a Gaussian kernel, and it reads as:
\begin{equation}
   W_L(y^{\prime}) = \int_{y_{\min}}^{y_{\max}} {\frac{\left| \hat{F}^+_{Y}(y)-\hat{F}^-_{Y}(y) \right|}{\min \left[ \hat{F}^0_{Y}(y),1-\hat{F}^0_{Y}(y) \right]}} e^{-\frac{(y-y^{\prime})^2}{2(\sigma(y^\prime))^2}} dy,
\end{equation}
where the  kernel parameters $\sigma(y^{\prime})$ is obtained by solving:
\begin{equation}
\label{eq_4}
\sigma(y^{\prime}) =  \{  { \sigma_{\hat{\mathcal{M}}}(\boldsymbol{x^\prime}) \mid}  \boldsymbol{x}^\prime =  \arg \underset{\boldsymbol{x} \in S}{\min} |y^\prime- \hat{\mathcal{M}}(\boldsymbol{x}) |   \}.
\end{equation}
Therefore, the threshold $y^*$ is optimized by solving:
\begin{equation}
\label{eq_2}
y^{*}=\arg \underset{y^{\prime}}{\max} \   \frac{1}{Z(y^{\prime})}  W_L(y^{\prime})     ,  y^{\prime} \in\left[y_{\min} , y_{\max }\right],
\end{equation}
where the normalization coefficient $Z(y^{\prime})$  is computed by 
 \begin{equation}
\label{eq_3}
Z(y^{\prime}) = \sqrt{2\pi}\sigma(y^{\prime})\left(   \Phi\left(\frac{y_{\max}-y^\prime}{\sigma(y^{\prime})} \right) - \Phi\left(\frac{y_{\min}-y^\prime}{\sigma(y^{\prime})}\right)                    \right),
\end{equation}
where $\Phi(\cdot)$ denotes the CDF of the standard Gaussian distribution. In the second step, we adopt the reliability-based learning function $U$ proposed in \cite{Echard2011} and it reads as:
\begin{equation}
\label{eq_5}
L_\text{t}(\boldsymbol{x}) =   \frac{   \lvert y^{*} - \hat{\mathcal{M}}(\boldsymbol{x}) \rvert  }{\sigma_{\hat{\mathcal{M}}}(\boldsymbol{x})} .
\end{equation}

\subsection{Gradient-based learning function}          
The gradient-based learning function \cite{mo2017taylor} is formulated by compromising global explorations and local exploitation:
\begin{equation}
\label{eq_6}
\begin{aligned}
L_{\text{g}}(\boldsymbol{x}) & =   L^\text{G}(\boldsymbol{x}) +\omega(\boldsymbol{x}) \times L^\text{L}(\boldsymbol{x}) \\
& = \frac{D_{\min}(\boldsymbol{x})}{\underset{\boldsymbol{x}\in S}{\operatorname{max}} D_{\min}(\boldsymbol{x}) }  + \omega(\boldsymbol{x}) \times \frac{  \hat{\mathcal{M}}^{{nl}}(\boldsymbol{x}|\mathcal{X}_{\min}(\boldsymbol{x}) )  }   {   \underset{  \boldsymbol{x} \in S  }{\operatorname{max}} \hat{\mathcal{M}}^{{nl}}(\boldsymbol{x}|\mathcal{X}_{\min}(\boldsymbol{x}) ) } ,
\end{aligned}
\end{equation}
where $ L^\text{G}(\boldsymbol{x})$  and  $L^\text{L}(\boldsymbol{x})$ score the benefits of adding $\boldsymbol{x}$ into the DoE from global and local search, respectively. Both these two scores are normalized by their corresponding highest score of the point in population $S$. $\omega(\boldsymbol{x})\in [0,1] $ is a weighting function that balances the exploration and exploitation, defined as:
\begin{equation}
\label{eq_7}
\omega(\boldsymbol{x}) = 1 - \frac{d(\boldsymbol{x}, \mathcal{X}_{\min}(\boldsymbol{x}))}{L_{\max}} ,
\end{equation}
where the normalized factor $L_{\max}$ denotes the maximum Euclidean distance of  points in the input DoE $\mathcal{X}$. $d(\cdot,\cdot)$ is the Euclidean distance measure, and $\mathcal{X}_{\min}(\boldsymbol{x}) \in \mathcal{X}$ is the point in the DoE with the minimum distance to points $\boldsymbol{x} \in S$. 

The global exploration searches the sparse DoE region and favors the points with larger values of $D_{\min}(\boldsymbol{x}) = d(\boldsymbol{x}, \mathcal{X}_{\min}(\boldsymbol{x}))$, i.e., the  points with maximum-minimum distances from $\boldsymbol{x} \in S$ to $\boldsymbol{x^{\prime}} \in \mathcal{X}$. The local exploitation targets  the region with strong nonlinearity via the Taylor series expansion of the surrogate model at point $\mathcal{X}_{\min}(\boldsymbol{x})$. Specifically, the surrogate prediction is decomposed as two terms:
\begin{equation}
\label{eq_9}
 \hat{\mathcal{M}}(\boldsymbol{x} ) =  \hat{\mathcal{M}}^{{l}}(\boldsymbol{x}|\mathcal{X}_{\min}(\boldsymbol{x}) )  +  \hat{\mathcal{M}}^{nl}(\boldsymbol{x}|\mathcal{X}_{\min}(\boldsymbol{x}) ),
\end{equation}
where the first term represents the linearly first-order expansion and the second term is the non-linearly high-order residual. The second term captures the extent of nonlinear extrapolation beyond the expansion point $\mathcal{X}_{\min}(\boldsymbol{x})$.  A larger contribution from this term to the total response indicates lower confidence in prediction accuracy. Therefore, the nonlinear residuals serve as an indicator used for locally refining the surrogates. 

\bibliographystyle{elsarticle-num}
\bibliography{MyRefs}